\PassOptionsToPackage{svgnames}{xcolor}
\documentclass[acmsmall,authorversion]{acmart}

\usepackage{adjustbox}
\usepackage{multirow}
\usepackage{makecell}

\usepackage{amsmath,amssymb,amsfonts}

\usepackage{graphicx}
\usepackage{textcomp}
\usepackage{xcolor}
\usepackage{booktabs}
\usepackage{hyperref} 
\usepackage{url}
\usepackage{subcaption}

\usepackage[para]{footmisc}

\usepackage{tikz}
\usepackage{pgf-umlsd}
\usepackage{adjustbox}
\usetikzlibrary{fit,positioning,arrows}
\usetikzlibrary{tikzmark}
\usetikzlibrary{shapes}
\usetikzlibrary{decorations.pathreplacing,calligraphy}
\usetikzlibrary{shadings}

\makeatletter
\begingroup\endlinechar=-1\relax
       \everyeof{\noexpand}%
       \edef\x{\endgroup\def\noexpand\homepath{%
         \@@input|"kpsewhich --var-value=HOME" }}\x
\makeatother
\def\overleafhome{/tmp}

\ifx\homepath\overleafhome
  \usepackage[frozencache,cachedir=.]{minted}
\else
  \usepackage[frozencache,cachedir=.]{minted}
\fi

\usepackage{import}
\ifx\homepath\overleafhome
  
\usepackage{etoolbox}

\newtoggle{AlgsNoEnd}

\providebool{tr}
\providebool{conf}

\usepackage[utf8]{inputenc}
\newcommand{\code}[1]{\texttt{#1}}
\usepackage[binary-units=true]{siunitx}
\usepackage{xspace}
\usepackage[svgnames]{xcolor}

\usepackage{collab}

\usepackage{soul}
\soulregister\cite7
\soulregister\autoref7
\soulregister\code7
\soulregister\toolname7
\soulregister\ref7
\soulregister\pageref7
\usepackage{hhline}
\definecolor{lightyellow}{RGB}{250, 250, 180}
\definecolor{HLYELLOW}{RGB}{240, 127, 0}
\definecolor{hlyellow}{RGB}{240, 127, 0}
\sethlcolor{lightyellow}

\usepackage{algorithm}
\iftoggle{AlgsNoEnd} {
  \usepackage[noend]{algpseudocode}
}{
  \usepackage{algpseudocode}
}
\usepackage{pifont}
\usepackage{amsmath}

\algdef{SE}[FOREACHIN]{ForEachIn}{EndForEachIn}[2]{\algorithmicfor\ \textbf{each}\ #1\ \textbf{in}\ #2\ \textbf{do}}{\algorithmicend\ \algorithmicfor}%
\algdef{SE}[FOREACHP]{ForEachP}{EndForEachP}[1]{\algorithmicfor\ \textbf{each}\ #1\ \textbf{in parallel do}}{\algorithmicend\ \algorithmicfor}%
\algdef{SE}[FOREACHINP]{ForEachInP}{EndForEachInP}[2]{\algorithmicfor\ \textbf{each}\ #1\ \textbf{in}\ #2\ \textbf{in parallel do}}{\algorithmicend\ \algorithmicfor}%
\algdef{SE}[FORMOD]{ForMod}{EndForMod}[3]{\algorithmicfor\ #1\ \textbf{from}\ #2\ \textbf{to}\ #3\ \textbf{do}}{\algorithmicend\ \algorithmicfor}%
\algdef{SE}[FORMOD]{ForModS}{EndForModS}[4]{\algorithmicfor\ #1\ \textbf{from}\ #2\ \textbf{to}\ #3\ \textbf{step}\ #4\ \textbf{do}}{\algorithmicend\ \algorithmicfor}%
\algdef{SE}[FORMODP]{ForModPS}{EndForModPS}[3]{\algorithmicfor\ #1\ \textbf{from}\ #2\ \textbf{to}\ #3\ \textbf{in parallel do}}{\algorithmicend\ \algorithmicfor}%
\algdef{SE}[FORMODP]{ForModP}{EndForModP}[4]{\algorithmicfor\ #1\ \textbf{from}\ #2\ \textbf{to}\ #3\ \textbf{step} #4\ \textbf{in parallel do}}{\algorithmicend\ \algorithmicfor}%

\iftoggle{AlgsNoEnd} {
  \algtext*{EndFunction}
  \algtext*{EndIf}
  \algtext*{EndFor}
  \algtext*{EndForEachIn}
  \algtext*{EndForEachP}
  \algtext*{EndForEachInP}
  \algtext*{EndForModP}
  \algtext*{EndForModPS}
  \algtext*{EndForModS}
}{}

\algdef{SE}[VARIABLES]{Variables}{EndVariables}{\algorithmicvariables}
  {\algorithmicend\ \algorithmicvariables}
\algnewcommand{\algorithmicvariables}{\textbf{global}}
\algnewcommand{\LineComment}[1]{\State \(\triangleright\) #1}
\algnewcommand{\And}{\textbf{and}\xspace}

\usepackage{tikz}
\usepackage{pifont}
\usetikzlibrary{shapes}

\usepackage{float}
\usepackage{listings}
\newfloat{codeblock}{H}{myc}
\definecolor{darkblue}{rgb}{0,0,.6}
\definecolor{darkred}{rgb}{.6,0,0}
\definecolor{darkgreen}{rgb}{0,.5,0}
\definecolor{red}{rgb}{.98,0,0}
\definecolor{gray}{rgb}{.6,.6,.6}
\definecolor{newgreen}{RGB}{169,209,142}
\definecolor{newpurple}{RGB}{237,134,254}
\definecolor{neworange}{RGB}{244,177,131}
\definecolor{newyellow}{RGB}{255,217,102}
\else

\fi

\newcommand{\toolname}{\emph{Cppless}\xspace}
\collabAuthor{copik}{blue}{MC}
\collabAuthor{lex}{green}{Lex}
\collabAuthor{htor}{red}{htor}
\collabAuthor{lukas}{black}{Lukas}

\definecolor{gold}{RGB}{153,102,51}
\definecolor{gray}{RGB}{118,113,113}
\definecolor{bg}{RGB}{238, 238, 238}

\usepackage{xcolor}
\usepackage{soul}
\usepackage{marginnote}

\setcopyright{cc}
\setcctype{by-sa}
\acmJournal{TACO}
\acmYear{2025} \acmVolume{1} \acmNumber{1} \acmArticle{1} \acmMonth{1} \acmPrice{}\acmDOI{10.1145/3747841}

\begin{document}

\title{Cppless: Single-Source and High-Performance Serverless Programming in C++}

\author{Marcin Copik}
\affiliation{%
	\institution{ETH Zurich}
	\country{Switzerland}
}
\email{marcin.copik@inf.ethz.ch}

\author{Lukas Möller}
\affiliation{%
	\institution{ETH Zurich}
	\country{Switzerland}
}
\email{mail@lukas-moeller.ch}

\author{Alexandru Calotoiu}
\affiliation{%
	\institution{ETH Zurich}
	\country{Switzerland}
}
\email{acalotoiu@inf.ethz.ch}

\author{Torsten Hoefler}
\affiliation{%
	\institution{ETH Zurich}
	\country{Switzerland}
}
\email{torsten.hoefler@inf.ethz.ch}

\begin{abstract}
	The rise of serverless computing introduced a new class of scalable, elastic and widely available
	parallel workers in the cloud.
	Many systems and applications benefit from offloading computations and parallel tasks to dynamically
	allocated resources.
	However, the developers of C++ applications find it difficult to integrate functions due to
	complex deployment, lack of compatibility between client and cloud environments, and loosely
	typed input and output data.
	To enable single-source and efficient serverless acceleration in C++, we introduce \toolname{},
	an end-to-end framework for implementing remote functions which handles the creation,
	deployment, and invocation of serverless functions.
	\toolname{} is built on top of LLVM and requires only two compiler extensions to automatically extract C++ function objects and deploy them to the cloud.
	We demonstrate that offloading parallel computations, such as from a C++ application to serverless workers,
	can provide up to 59x speedup with minimal cost increase while requiring only minor code modifications.
\end{abstract}



\maketitle

\section{Introduction}

Serverless functions have taken cloud systems by storm.
Stateless, short-lived, and isolated functions execute on dynamically allocated cloud resources, and the programming model of
Function-as-a-Service (FaaS) hides the software and hardware stacks of the cloud from the user.
Functions offer a highly scalable and elastic offloading of computations to dynamically allocated parallel workers,
with up to 6000 new function containers in a minute on commercial cloud platforms~\cite{awsLambdaScaling}.
While serverless has initially gained popularity in web development and API integration, functions have been recently used for parallel and compute-intensive tasks such as data analytics, machine learning training, compilation, and high-performance computing~\cite{jiang2021towards,10.1145/3267809.3267815,201559,227653,copik2021rfaas,10.5555/3277355.3277444}.

In the pay-as-you-go system of FaaS, users are charged for each millisecond of active computation in a function.
In a traditional deployment to virtual machines in Infrastructure-as-a-Service (IaaS), optimizations primarily
improve the responsiveness and throughput of services.
However, these improvements might not immediately lead to decreased costs when rescaling the deployment is not feasible.
On the other hand, serverless functions immediately benefit from optimized code and shortened runtime as every millisecond saved directly decreases the cost of running an application in the cloud.
Shaving off seconds from functions is crucial in FaaS to provide lower costs than a persistent IaaS deployment for large and complex workloads \citep{copik2021sebs,copik2022faaskeeper}.
Hence, it is important that services are implemented on efficient backends and gain from the performance of compiled languages.

\begin{figure}[t]
	\centering
	\includegraphics[width=1\textwidth]{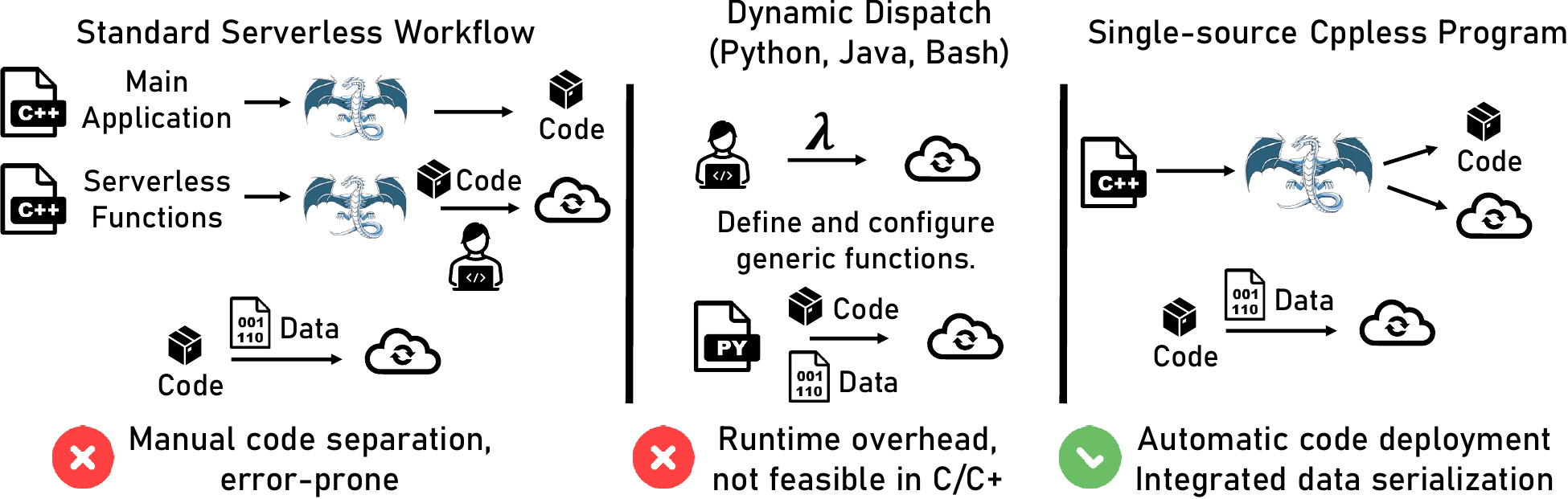}
	\caption{Compilation in \toolname{}: single-source programming with automatic deployment at compilation time.}
	\label{fig:posterchild}
\end{figure}

Even though serverless has been shifting to larger and more compute-intensive workloads,
it is still dominated by languages such as JavaScript and Python.
These languages make it particularly easy to implement serverless functions, thanks to embedded support for serialization, JSON, HTTP request handling, and an easy deployment mode with function code extraction at runtime.
However, achieving high performance in these high-level and dynamic languages is difficult and requires programmers to offload hot spots to lower-level languages with the help of native extensions.
Interpreted languages are particularly affected by high costs of cold startups (Section~\ref{sec:eval-pi-cold}).
Furthermore, many parallel and high-performance libraries and frameworks are already implemented in C/C++ and could be used for parallel computations in the cloud.
Unfortunately, the drive toward a high-performance serverless cloud is hindered by the complex deployment model
and lack of native integration of serverless functions into C++ applications.

Serverless C++ functions require developers to split functions from the main application and compile them separately,
deploy to the cloud using interfaces that are not standardized and differ for each cloud platform,
and execute through a vendor-specific REST and RPC APIs.
While managed languages can use the bytecode and runtime introspection mechanisms to automatically
extract function code from an application,
existing C++ language capabilities are not sufficient for such a task.
This problem is aggravated by the lack of compatibility between client and cloud software and hardware environments, and
the conversion of statically typed data structures from and into the loosely typed JSON format (Section~\ref{sec:background_cppless}).
This results in a convoluted process and a high entry barrier for serverless functions in high-performance applications.
To benefit from serverless acceleration, parallel C++ applications need a framework that keeps the application and function code together to achieve high productivity,
while avoiding code bloat and cloud vendor lock-in.

We resolve the aforementioned limitations by introducing \toolname{}, a single-source programming model and an end-to-end compiler for serverless functions in C++ (Figure~\ref{fig:posterchild}).
\toolname{} accelerates parallel C++ applications by shifting compute-intensive tasks to serverless functions, which are automatically created, deployed, and invoked by the framework.
Similarly to GPU programming frameworks like CUDA and SYCL, which embed kernel code within C++ applications, we include remote serverless functions as regular function objects and C++ lambda functions.
In the example of a Ray-Tracer application, using serverless with \toolname{} requires only 7 lines of offloading code
and 18 lines of serialization instructions more than a multithreaded implementation,
and provides up to 8.26x speedup against local computation on a virtual machine with 16 vCPUs.
Obtaining the result with functions costs less than 0.0027\$ and is delivered in one second, while a virtual machine could take several minutes to boot,
and keeping online even a small general-purpose virtual machine \code{t2.medium} with two virtual CPUs costs 0.0536\$ per hour - almost twenty times more.

\toolname{} achieves these performance, cost, and productivity results by combining the serverless and non-serverless program parts in a single source code.
To that end, we introduce the concept of \textbf{alternative entry points} to redirect the compilation of a single translation unit into multiple targets.
Once the compiler detects a serverless function code inside a C++ translation unit, it splits the compilation path for the selected function to generate a separate executable (Section~\ref{sec:compilation}).
To dispatch C++ lambda functions to the cloud, we extended the compiler with two additional features: \textbf{reflection} to access capture variables and \textbf{unique naming scheme} for these anonymous functions.
\toolname{} automatically deploys alternative entry points to the cloud as serverless functions.
A serialization library helps pack input data into binary format, and the framework encapsulates the vendor-specific process of deploying and invoking functions.

The \toolname{} compiler is built on top of the LLVM framework and requires only a few modifications to the \code{clang} codebase (Section~\ref{sec:implementation}).
To seamlessly integrate serverless functions into applications, we implemented a high-level and templated user interface, which internally uses the language extensions added to support single-source programming.
With just a few lines of code added to the application, users can offload native C++ computations to elastic and scalable serverless functions while retaining \textbf{the same compilation workflow}.
With a set of microbenchmarks and parallel applications (Section~\ref{sec:evaluation}), we demonstrate that \toolname{} enables efficient parallel offloading to the serverless cloud without compromising the productivity and safety of C++ programming.

In this paper, we make the following contributions:
\begin{itemize}
	\item Serverless programming model for efficient offloading computations in C++ to the cloud.
	\item A C++ toolchain extended with alternative entry points and serializable lambda functions, allowing users for straightforward embedding of serverless functions into their applications. The toolchain is available on an open-source license,\footnote{\url{https://github.com/spcl/cppless}} and is accompanied by a paper artifact.\footnote{\url{https://doi.org/10.5281/zenodo.15778386}}
	\item A C++ framework providing high-level abstractions that hides the complexity of cloud provider APIs, demonstrated with the example of AWS Lambda.
\end{itemize}

\section{Why \toolname{}?}

\begin{figure}[t]
	\centering
	\includegraphics[width=1.0\linewidth]{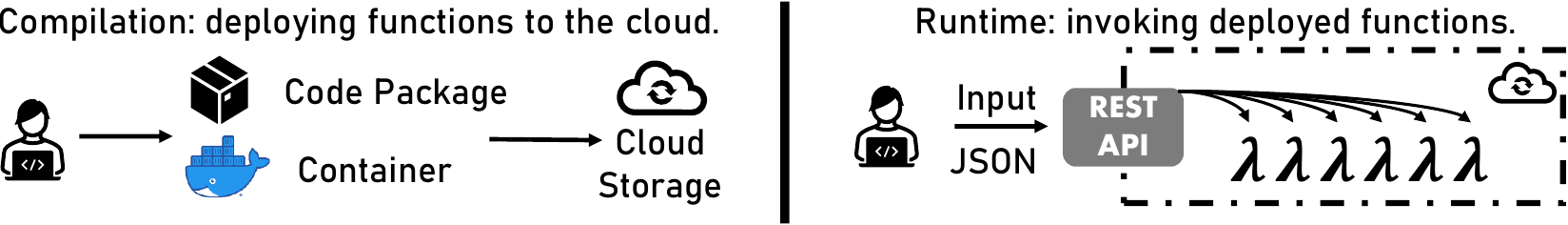}
	\caption{\textbf{Development workflow in serverless}:
		functions are shipped to the cloud during deployment.
		At runtime, the main applications invoke previously created functions.
	}
	\label{fig:faas_overview}
\end{figure}

Serverless functions provide high scalability but introduce a division between functions shipped
to the cloud and the main application code (Figure~\ref{fig:faas_overview}).
While several frameworks have been developed to integrate functions and applications using them,
they targeted high-level, interpreted, and dynamically-typed languages (Section~\ref{sec:related_work}).
In \toolname{}, we attempt to achieve the same goal for C++:
a standard-compliant interface for single-source serverless programming.
To that end, we first examine the low popularity of C++ in serverless.
Then, we identify the unique challenges of bringing the statically typed and compiled C++
to the serverless world.

\subsection{Programming Languages in Serverless}
\label{sec:background:lang}
Serverless computing has been dominated by dynamically typed and interpreted
languages like Python and JavaScript,
covering 70\% of serverless invocations~\cite{stateServerless}
and 85\% of open-source serverless applications~\cite{9543531}.
The serverless landscape differs fundamentally from the rest of the programming world,
as shown in the most recent TIOBE index~\cite{tiobeIndex}.
There, Python and JavaScript achieved
the first and sixth positions with ratings of 16.33\% and 3.01\%, respectively.
On the other hand, C and C++ occupy the second and third place, respectively,
achieving a cumulative rating of 19.5\%.
This difference is not caused by a lack of C++ applications that can benefit from serverless computing, as shown in the examples of Monte-Carlo simulations~\cite{copik2021rfaas,copik2024disagg}
and distributed scientific analyses for large data processing~\cite{9826036}.
Instead, we hypothesize that the low adoption is caused by insufficient support for integrating functions into C++ codebases and seamlessly deploying to the cloud.

\textbf{Productivity}
FaaS is a new programming model that requires developers to adapt their codebases and
toolchains to cloud deployment.
%
A taxonomy of challenges experienced by serverless developers
contains two categories directly related to \toolname{}:
serverless application implementation and deployment~\cite{10.1145/3468264.3468558},
When considering the implementation of serverless applications, developers particularly
struggle with integrating the deployment package and invoking functions through the cloud API.
Deploying to FaaS causes issues in building a code package according to the
size restrictions and expected format and deploying it to the cloud with available CLI tools.
Thus, a productive programming framework needs to provide solutions for both categories
to ease the burden and allow developers to focus
on their application's logic and performance.

\subsection{C++ Challenges in Serverless}
\label{sec:background_cppless}
Integrating serverless functions into C++ applications is difficult because of
three major differences and challenges that are resolved in the \toolname{} framework.

\textbf{Challenge \#1: Compile-Time Dispatch}
Interpreted languages such as Python and Java permit extracting function code and embedding it with the input data into a POST request,
allowing for dynamic function dispatch at runtime.
On the other hand, C++ requires that the function code is compiled ahead of time, and it lacks runtime introspection and reflection mechanisms that could reliably and
efficiently discover all dependencies of a selected function.
Shipping all linked libraries would quickly lead to transmitting dozens and hundreds of megabytes, creating major performance overheads.
Instead, the C++ program must be restructured to allow for separate compilation of the serverless functions and extended with serialization and invocation interface.
Then, the results of the compilation should be uploaded to the cloud once the compilation is finished.
This way, users can execute their C++ applications immediately after finishing the build process.

\textbf{Challenge \#2: Server Environment}
\emph{Serverless} functions still execute on a \emph{server}, which can be easily hidden in high-level and interpreted languages.
However, this is not the case for compiled languages such as C++, where both the underlying architecture and ABI compatibility are of concern.
For example, the user code might be running on an ARM notebook and link against the \code{libc++} and \code{libc} standard libraries, while the function code will
execute in a Linux container on an x86 server, with \code{libstdc++} and \code{musl} available as implementations of the C++ and C standard libraries, respectively.
Thus, more is needed than just to split the function code into a separate shared library; the application can no longer be compiled in a single environment with the same configuration.
Languages such as C++ require a new programming model that will handle function code as a separate entity,
while hiding this complexity from the user behind a uniform interface.

\textbf{Challenge \#3: Static Typing}
Cloud platforms accept function input in a JSON format through a RESTful interface, which requires a dedicated conversion from strongly typed data structures.
This is in contrast to microservices built on top of RPC communication, which hides the network transport of input arguments from the developer with the help of frameworks like Protobuf in gRPC and Transport in Thrift.
Languages like Python and JavaScript come with native JSON support and serialization of objects to this format; C++ has no standard (de)serialization procedures.

C++ ensures type safety through compile-time verification.
However, when function implementation is not single-source, and clients are connected to serverless environments only by loosely typed JSON, this compile-time safety is lost.
When separated implementations of the invoking application and \emph{serverless} function are joined only by the loosely typed JSON, type verification becomes the responsibility of the serialization runtime.
JSON provides a limited possibility of encoding types, e.g., it is not possible to distinguish between integers of different sizes.
This forces developers to implement manual JSON schemas, complex error handling, or adopt specialized serialization libraries that can properly encode diverse C++ types.
This approach effectively shifts type checking from compilation to runtime, increasing the time and size overhead.

\section{Designing a C++ Compiler for Serverless Functions}

\label{sec:compilation}

We begin by introducing the \toolname{} interface from a user perspective (Section~\ref{sec:design_example}).
\toolname{} is designed as a group of C++ language extensions (Section~\ref{sec:design_extensions}),
implemented in the LLVM compiler (Section~\ref{sec:implementation}).
The extended compiler is combined with deployment tools to
provide a seamless, single-step, and productive process of deploying C++ software to serverless
(Section~\ref{sec:design_sending_cloud}).
For details on the design and implementation of \toolname{}, we refer readers to the work of Möller~\cite{thesislukas}.

\begin{figure}
	\begin{minted}[linenos,fontsize=\footnotesize,escapeinside=??]{cpp}
double pi_estimate(int n); // Implementation of the Monte Carlo estimation

double compute_pi() {
  const int n = 100000000, np = 128;

  cppless::aws_dispatcher dispatcher;?\tikzmark{here}?
  auto aws = dispatcher.create_instance();
  using config = lambda::config<cppless::lambda::with_memory<512>>;

  std::vector<double> r(np); 
  // Define lambda function offloaded to serverless cloud
  auto fn = [=] { return pi_estimate(n / np); };
  for (auto& result : r) {
    // Invoke remote cloud functions. Returned values are written to the vector
    cppless::dispatch<config>(aws, fn, result);
  }
  cppless::wait(aws, np);  // Wait for all invocations to finish

  return std::reduce(r.begin(), r.end()) / np;
}
\end{minted}
	\caption{Offloading parallel PI computation to AWS Lambda with \toolname{}.
		The program launches 128 functions in parallel to process 100 million samples.
	}
	\label{lst:example_pi}
\end{figure}

\subsection{Parallel Computing with \toolname{}}
\label{sec:design_example}

We use the classic example of parallelizing Pi estimation (Figure~\ref{lst:example_pi}) to demonstrate how users can implement serverless
functions with \toolname{}, while not making their code dependent on additional, third-party cloud frameworks
and keeping the function code united with the main application.
In order to execute the computations on serverless functions, the user creates an instance of a dispatcher configured for the selected cloud system (line 7).
In this case, we consider executing C++ functions as AWS Lambda functions.
Within a dispatcher instance, all invocation requests share the same network resources, such as HTTP sessions, and each request is assigned a unique local identification.
To invoke the function concurrently across 128 instances of an AWS Lambda function, the user calls the \code{cppless::dispatch} function (line 15),
which will order the \toolname{} compiler to turn the user C++ lambda function (line 12) into an AWS Lambda function in the cloud.
Users can optionally configure resources assigned to the cloud function, such as memory and temporary storage (line 8).
The compiler will convert the configuration to metadata and attach it to the generated function code.
As a final compilation step, \toolname{} will upload the generated function code to the cloud and create a function using the cloud provider APIs.

At runtime, the call to \code{cppless::dispatch} triggers an asynchronous function invocation by sending an HTTP request.
The values of \code{n} and \code{np} captured in the C++ lambda function are serialized (line 12),
and \toolname{} uses internally the C++ library \code{cereal}~\cite{cereal} for that task.
The dispatcher selects the AWS Lambda function to be invoked through the unique type identification generated by the \toolname{} compiler.
The third parameter of this function specifies where the result should be stored.
Then, \texttt{cppless::wait} blocks until all invocations are finished (line 17).
The \toolname{} runtime deserializes the output of function execution, which can be read directly by the user and merged to the final result (line 19).
In addition to the return value, each invocation returns the cloud-assigned unique invocation identifier and a flag set to true if the invocation is cold.

\subsection{C++ Language Extensions}
\label{sec:design_extensions}
Alternative entry points are the most crucial part in the \toolname{} compilation flow: they enable a single compilation unit to expose multiple
entry points, letting us generate code for many serverless functions from a single C++ translation unit.
Complemented with the second extension of lambda reflection,
the compiler can now dispatch C++ function calls to the cloud.

\begin{figure}[t]
	\begin{minted}[fontsize=\footnotesize]{cpp}
  template <class Func>
  struct ProcessBridge {
    int operator()() {
      auto alt_entry_name = gen_id<Func>(); // Spawn execution of the alternative entry point.
      return spawn(alt_entry_name);
    }
    __attribute((entry)) int main(int, char**) {
      Func func;
      deserialize(read_http_request(), func);
      respond_http(func());
    }
  };
\end{minted}
	\caption{\code{ProcessBridge} presents a simplified example of connecting a user-defined
		function object with an invocation of a remote serverless function.}
	\label{lst:processbridge}
\end{figure}

\paragraph{Alternative Entry Points}
\toolname{} exports the code of serverless functions to a separate compilation path
by defining \textit{alternative entry points}.
Many programming languages define the concept of an entry point, which is a function executed when the program is started. 
From the entry point, the control flow can diverge and is governed by the language's semantics.
In C-family languages, this function is usually called \code{main}, which is automatically
called at the begin of execution.\footnote{The actual start is typically done through the \code{\_start} function which is
	the entry point.}

From a user perspective, alternative entry points are annotations added to a function declaration.
Adding this annotation affects the compilation process by creating a separate executable
or library where the
the main function is replaced with the body of the alternative entry point function.
However, users are not expected to directly use the alternative entry points; nor
do they need to be aware of their existence,
as this compiler feature is hidden behind the \code{cppless} interface.
Function objects are used in composition with entry points to implement
\emph{bridge classes}, which use template programming to provide an interface to an
alternative entry point that they define (Figure~\ref{lst:processbridge}).
This process can be used to model and deploy serverless functions:
The client-side representation of a serverless function is an instance of the bridge class,
the instantiation of this bridge class also automatically registers an alternative entry point
that the runtime can use to create an invocation.

\begin{figure}[t]
	\begin{minted}[fontsize=\footnotesize]{cpp}
template<typename Func>
void serialize_lambda(Func && lambda) {
  double arg1 = 42; std::string arg2 = "capture";
  auto lambda = [arg1, arg2]() { std::cout << arg1 << " " << arg2 << std::endl; };

  // The number of captured values is 2
  constexpr int capture_count = decltype(lambda)::capture_count();
  // Returns 42, type of variable: double
  auto captured_arg1 = lambda.capture<0>();

  lambda.capture<0>() = 43; 
  // Will print "43 capture"
  lambda();
}
\end{minted}
	\caption{Example demonstrating our language extension: lambda reflection.
		\toolname{} allows users and libraries to inspect the capture, their types, and values.
	}
	\label{lst:example_lambda_captures}
\end{figure}

\paragraph{Function Serialization}
In \toolname{}, serverless functions are defined as function objects,
including C++ lambda functions, which define their own custom type.
To seamlessly dispatch function invocations, \toolname{} needs to serialize function arguments.
In the case of lambda functions, serialization also includes the state variables captured in the closure.
To that end, we introduce a second extension:
a compile-time, constexpr-compatible reflection mechanism for lambda functions
(Figure~\ref{lst:example_lambda_captures}).
The reflection exposes direct accessors to the hidden unnamed capture members.
The template member function \code{capture} returns an l-value reference that can be used both to read and to write the individual unnamed capture members.
Thus, the host serialization can read the type and value of a state variable, while the serverless function can update the state based on the deserialized input data.
Thus, \toolname{} can serialize functions if all variables captured in the function are serializable.

\paragraph{Why Language Extensions?}
Similarly to single-source GPU frameworks, \toolname{} compiles application code into
multiple objects, which cannot be implemented with standard C++ features.
Separating entry points allows us to optimize functions by targeting different architectures and adapting compilation options.
Furthermore, a solution using a single executable with runtime dispatch would require additional code changes to adapt \code{main} function and assist discovery of serverless functions, increasing the burden on the user and making the tool less convenient.

Serializing lambda functions is not possible in C++ due to lack of reflection.
Future versions of C++ might include reflection support~\cite{cppReflection};
\toolname{} supports offloading lambda functions today and can adopt standard-compliant solutions once they become available.
Furthermore, the internal storage of lambda capture is not fully specified, and it might not include variables captured by reference.
In practice, compilers implement reference captures by adding pointers or references to lambda's storage~\cite{cppLambdaClosure}, and \toolname{}
makes them explicitly available to users and libraries.

\begin{figure}[t]
	\centering
	\includegraphics[width=\textwidth]{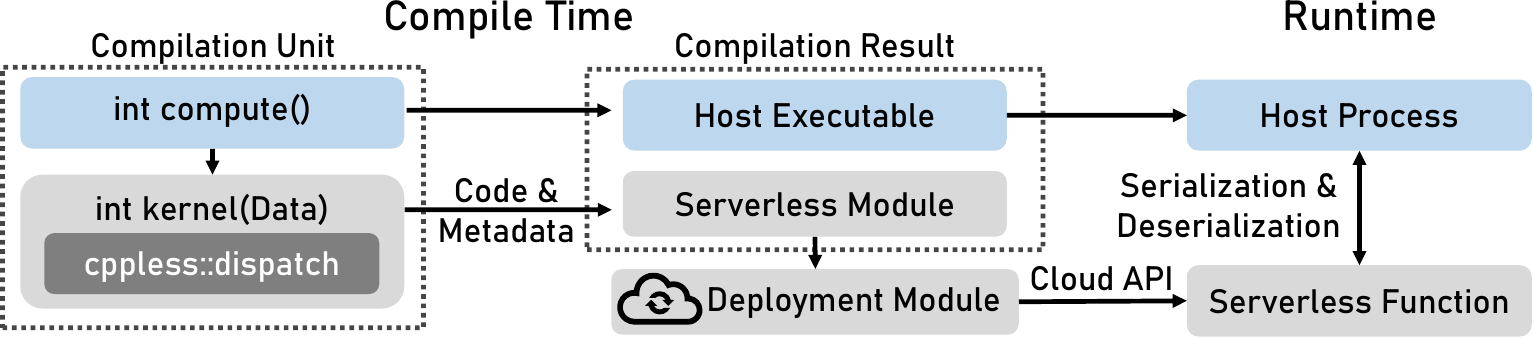}
	\caption{\toolname{} exports selected functions as additional compilation targets, deploys them to the cloud, and provides invocation and serialization mechanisms at runtime.}
	\label{fig:cppless_overview}
\end{figure}

\subsection{Sending Code to the Cloud}
\label{sec:design_sending_cloud}
In \toolname{}, the user C++ code is separated into the single host process and potentially
multiple serverless functions, which are deployed to the cloud and invoked remotely (Figure~\ref{fig:cppless_overview}).
First, each serializable function object marked for \emph{FaaS-ification} is wrapped with
a bridge class to define an alternative entry point.
Free functions, which are not bound to a particular function object, can be serialized by
wrapping them with lambda functions.
The serialization of the input and output of the function is implemented in a runtime library
and relies on the \code{cereal} library,
including support for serialization of many types from the C++ Standard Template Library,
such as \code{std::string} and \code{std::vector}.
The user only has to manually add serialization for custom types, which is necessary as
C++ objects cannot be serialized in a unique, cross-platform way.

Once the user application attempts to invoke a serverless function, the \toolname{}
library will call the serialization methods for each input argument and identify
the external cloud function through the type name.
In the function code, the bridge class code is responsible for deserializing the
input arguments and calling the original function object.
The bridge class entry point interacts with the cloud provider
environment, e.g., reading the arguments using the provided function API and writing function results.
The return value is serialized and deserialized in the same fashion.

\section{\toolname{} Implementation}

\label{sec:implementation}

The \toolname{} runtime library builds on top of the new language extensions and hides all cloud vendor-specific cloud interfaces (Section~\ref{sec:implementation_dispatcher}).
In addition to alternative entry points and lambda reflection, we add a unique name for all lambda functions with adjusted mangling rules (Section~\ref{sec:implementation_lambda}).
Internally, all language extensions are implemented
directly in the LLVM and Clang compiler~\cite{LLVM:CGO04} (Section~\ref{sec:implementation_llvm}),
The interactions with multiple compiler passes and cloud deployment is hidden from the user through extensions to the build system (Section~\ref{sec:implementation_pipeline}).
Finally, we discuss the limitations of the current \toolname{} implementation and propose solutions to overcome them (Section~\ref{sec:implementation_limitations}).

\begin{figure}
	\begin{minted}[fontsize=\footnotesize]{json}
"entry_points": [{
    "original_function_name": "mangled_C++_function_name", "filename": "dispatcher_aws_alt_0",
    "user_meta": {
      "ephemeral_storage": 512, "memory": 1024, "timeout": 10,
      "identifier": "./examples/aws/dispatcher.cpp@..."
    }
}]
\end{minted}
	\caption{
		Alternative entry point metadata, as generated by the \toolname{} compiler.
	}
	\label{lst:example_lambda_metadata}
\end{figure}

\subsection{User Library with Dispatcher}
\label{sec:implementation_dispatcher}
\toolname{} implements a \emph{fork-join} style API built on top of a low-level dispatcher interface,
which is based on sending tasks in the form of serializable and identifiable function objects.
In addition to \emph{joining} on a single function result, dispatchers also offer a \emph{wait-any}
operation.
Dispatchers encapsulate an interface of a single cloud provider, allowing to easily switch between
different systems without requiring users to rewrite their applications;
user-code can be implemented as a C++ template with dispatcher type as a template parameter.
Dispatchers interact with the compilation pipeline through the metadata system
(Figure~\ref{lst:example_lambda_metadata}).
%
%
The metadata list makes it possible to define configuration options on a per-function level directly in the C++ application.

We implement two methods of generating HTTP requests to trigger serverless functions,
an HTTP/2-based implementation with \code{nghttp2}~\cite{nghttp2}, and an HTTP/1.1-based implementation
that uses the \code{Boost.Beast} library~\cite{beast}.
We implement two solutions as they have different trade-offs: HTTP/2 can achieve better performance when sending multiple requests simultaneously,
while the Boost-based implementation increases portability by complementing the HTTP/2-only \code{nghttp2}.

With nghttp2, we distribute invocation requests in a round-robin fashion to a group of connections to the AWS service.
Instead of invoking functions through their own and custom HTTPS addresses, connections are initialized to the AWS Lambda REST API, and we specify the function name in each REST request.
With this solution, we can reuse warm connections for consecutive invocations, support many concurrent
requests, and reduce the chance of head-of-line blocking issues.
On the other hand, the Boost-based implementation issues a TCP-backed HTTP request for each invocation.
All requests share the same instance of the Boost.Asio IO context to manage asynchronous requests.
However, separate requests mean that the number of concurrent invocations is limited by the space of file descriptors available to the user process.

\begin{figure}
	\begin{minted}[fontsize=\footnotesize]{cpp}
template<typename Func>
void dispatch_function(Func && lambda) {
  // Cppless Language Extension: Lambda Reflection
  constexpr int capture_count = decltype(lambda)::capture_count();
  auto first_capture = serialize(lambda.capture<0>());

  // Unique lambda names, backed by sycl-unique-stable-name but with inline namespaces disabled
  auto func_id = __builtin_unique_stable_name(decltype(lambda));
  invoke(func_id, first_capture);
}
\end{minted}
	\caption{Pseudocode of serverless dispatch: combining reflection,
		unique identification of lambda functions, and serialization library
		with user-supplied customizations.
	}
	\label{lst:example_lambda_dispatch}
\end{figure}

\subsection{Lambda Functions}
\label{sec:implementation_lambda}
To differentiate between many deployed serverless functions,  the compiler must create a connection between the function object code and the remote entry point.
However, in contrast to regular free functions, lambda functions in C++ are unnamed.
To solve this problem, we generate unique identification of all types and use these compile-time values to name alternative entry points (Figure~\ref{lst:example_lambda_dispatch}).
This functionality is backed by the Clang implementation of \code{builtin-sycl-unique-stable-name},
a feature added for to support the SYCL framework~\cite{10.1145/2791321.2791345}, which has a similar use-case.
The function generates a mangled type name, but uses a slightly modified Itanium C++ ABI mangling~\cite{itaniumabi} where we remove inlined namespaces from the mangling prefix.
This feature is controlled by an additional flag and is disabled by default to not affect other mangled names.
This increases compatibility when client and function code are compiled with different standard library implementations (Section~\ref{sec:background_cppless}).

\subsection{A Serverless Compiler in Clang}
\label{sec:implementation_llvm}
The changes to the Clang compiler are limited to 963 lines of code added and 192 lines removed.

\paragraph{Frontend}
The compiler frontend is altered to parse and validate the new annotation for alternative entry points and metadata.
As alternative entry points must not be directly called, issues arise where they
are not emitted into the LLVM module, especially with top-level declarations.
We ensure that methods annotated as alternative entry points are treated as if they are used to prevent dead-code optimizations, and we force the template instantiation when the entry point is present.
Clang's \code{used} annotation has a similar effect.
In templated contexts, this change ensures the function is emitted once the parent context is fully instantiated.

\begin{figure}[t]
	\centering
	\includegraphics[width=\textwidth]{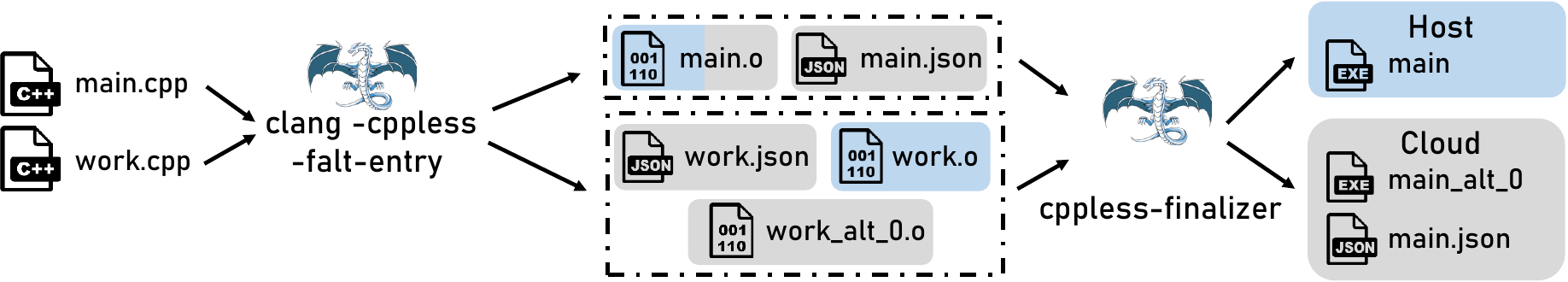}
	\caption{Compilation code flow of a project with an alternative entry point in \code{work.cpp}, compiled to
		\code{work\_alt\_0.o}. The CMake extension hides the multiple compilation steps and deployment to the cloud.}
	\label{fig:code_flow_overview}
\end{figure}

\paragraph{Code Generation}
We amend the code generation process in LLVM to work with alternative entry points.
We propagate the names of functions exported to the cloud to the subsequent compilation steps.
The metadata describing the function configuration is evaluated as a constant expression.
For each alternative entry point, we convert the metadata to a \code{std::string} instance available
to the compiled program and attach it to the corresponding LLVM function as an attribute.

\paragraph{LLVM Backend}
During the backend code generation, we propagate information about alternative entry points through the pipeline.
The corresponding LLVM function is annotated as an alternative entry point to ensure a separate treatment in the backend.
Once the CodeGen module generates the main LLVM module, it is cloned for each entry point found in the translation unit.
For each such module, we generate the code which results in a separate object file for each alternative entry.
Additionally, a binary is created for the host application.
At this point, we also create the manifest file, which stores configuration data of all entry points, including the user-supplied metadata, such as function resource configuration.

\paragraph{Linking}
The main Clang driver handles all linker invocations for specified object files and targets.
Build tools often utilize this driver due to its uniform interface,
which motivates building a modified linker driver that exposes the same interface and can be used by build systems.
We introduce a new tool \code{cppless-finalizer}, a cross-platform driver for Clang that can handle the linking step for multiple output files when alternative entry points are present.
\code{cppless-finalizer} accepts the same command line interface as Clang,
using the same Clang toolchains to support linking for different platforms.
Our tool reads manifest files from the compilation which describe alternative entry points, and links them using the original Clang driver.
The new driver produces one regular output file and additional files for each alternative entry point,
while merging the manifest files into a single result (Figure~\ref{fig:code_flow_overview}).

\subsection{Compilation Pipeline}
\label{sec:implementation_pipeline}
When using alternative entry points, \toolname{} produces an additional executable
for each serverless function (Figure~\ref{fig:code_flow_overview}).
Since the bridge class templates are instantiated lazily, the alternative entry points are
only generated for serializable function objects if they are indirectly instantiated from
non-templated contexts.
We implement a custom deployment tool in addition to the C++ compiler, which is responsible for
deploying the compiled function code using metadata stored in a compiler-emitted manifest file.
We propose a compilation flow consisting of (1) integration with the CMake build system,
(2) a deployment tool that uploads alternative entry point executables to the
cloud as serverless functions.
The former provides users with automatic invocation of all compilation steps,
while the latter encapsulates cloud-specific interfaces.

\begin{figure}
	\begin{minted}[fontsize=\footnotesize]{cmake}
# Create standard C++ compilation target
add_executable(parallel_pi parallel_pi.cpp)
# C++ target depends on our runtime
target_link_libraries(parallel_pi PRIVATE cppless::cppless)
# Enable processing of alternative entry points, and adds AWS C++ runtime to function
aws_lambda_target(parallel_pi)
# Deploys functions to the AWS cloud
aws_lambda_serverless_target(parallel_pi)
\end{minted}
	\caption{Integrating \toolname{} into the build system requires
		only minor adjustments to build targets.
	}
	\label{lst:example_pi_cmake}
\end{figure}

\paragraph{CMake Build Integration}
To generate and deploy serverless functions, users add \toolname{} to their
project build system (Figure~\ref{lst:example_pi_cmake}).
We implement CMake extensions that define specific build targets with support for compilation with \toolname{}.
The target adds a second compilation pass, adjusts compiler flags,
and invokes cloud deployment tools when code change is detected.

Each target with alternative entry points is compiled twice.
First, we conduct the compilation for the host system when alternative entry points are not emitted, and we resolve all dependencies and configuration details for the known host target.
Since this compilation process can use different compilation flags and linker targets than needed for the serverless functions, we create a different CMake configuration for the remote target.
Internally, separate build configurations are managed by using the \code{ExternalProject} functionality of CMake: we instantiate the project again, enable the \emph{serverless} setting,
and build only the selected target in the newly created project.
In this new project, we change compilation settings to invoke Clang in the mode that emits alternative entry points.
Furthermore, the project's nested instance can use different flags and link targets for the \emph{serverless} mode.
The user can provide a different CMake toolchain for the alternative mode to adapt code generation for the target environment of serverless functions.

These details are hidden from the user by exposing a function
\code{aws\_lambda\_serverless\_target} in CMake (Figure~\ref{lst:example_pi_cmake}).
The function effectively creates an additional cross-compiled target.
The user must decide the cloud environment at the configuration time by selecting a specific function in CMake.
The specialization to a specific FaaS platform happens at the compilation time because of differences between clouds in target architecture, environments, and deployment steps.

\paragraph{Cloud Deployment Script}
The deployment script encapsulates the complexity of managing cloud resources and
vendor-specific interfaces.
For each alternative entry point, a new serverless function is created, configured,
and deployed with the compiled code.
Function names are provided through the unique type identifiers generated by the compiler.

\paragraph{Cross-Compilation}
\toolname{} generates alternative entry points within a separate CMake project instance, enabling compilation with completely different toolchains.
We demonstrate this with a simple example of offloading two functions to the cloud,
which are cross-compiled from an x86 host to the ARM instance of AWS Lambda running on Graviton CPUs.
The implementation process involves extracting the ARM sysroot from official Amazon Linux AWS containers and configuring a custom CMake toolchain file specifying the ARM target triple and sysroot location.
All \toolname{} dependencies are built using this cross-compilation toolchain, while the packaging process is adapted to identify dynamic dependencies without relying on the native x86 \code{ldd} utility.

\subsection{Limitations}
\label{sec:implementation_limitations}
\paragraph{Serializable Function Objects}
Requiring tasks to be serializable functions permits using some constant function pointers, i.e., a pointer where the exact name of the called function is known at compile time.
Supplying function pointers does not fit the user-space design where the type of function objects is used as a basis for creating serverless functions.
Instead, a potential solution would be to add an implicit conversion from constant function pointers to captureless lambda function objects.
This would allow \toolname{} to treat function pointers as regular lambdas.

\paragraph{Entry Point Interface}
Since each alternative entry point is treated as a clone of the \emph{main} function,
it must use the same interface and be compiled into an executable.
In the example of AWS Lambda, this aligns with the design of the platform,
where users deploying C++ functions are expected to deploy an executable that
starts a server waiting for new requests and manually process them with the selected function.
However, other platforms might require different forms of deploying code, such
as shared libraries.
There, \toolname{} could generate shared libraries instead.

\paragraph{Selective Compilation}
\toolname{} currently treats all alternative entry points in the same way and generates full code for each one of them.
To decrease the size of alternative entry points, \toolname{} could incorporate techniques from established offloading frameworks that reduce the number of functions compiled, e.g., explicit annotations of offloaded functions like \texttt{\_\_device\_\_} in CUDA or \texttt{\#pragma omp declare target} in OpenMP.
Additionally, functions can also be determined automatically based on use in the explicitly offloaded code, as is the case for SYCL and OpenMP.

\paragraph{Compilation Time}
As multiple LLVM modules are produced and then lowered to the target language independently, the compilation time of a target increases: each target with alternative points is passed to the compiler twice, and the code generation time of the target increases linearly with the number of alternative entry points.
This limitation could be fixed by integrating the cloning process deeper with LLVM, which currently does not support such integration.
Thus, we limit ourselves to using the public LLVM APIs to improve the maintainability of \toolname{}.

\paragraph{Captured Variables}
\toolname{} supports passing arguments only by value, with modified data returned explicitly as the function result.
This design addresses potential complications in serverless environments: functions typically execute asynchronously and treating reference-captured variables as value-captured prevents issues with dangling references and memory synchronization.

In \toolname{}, the values of variables captured within C++ lambda expressions are transmitted as the function invocation payload.
This means we exclude non-local (global) variables since they are not included in the lambda's capture list.
If access to global variables becomes necessary, this limitation could potentially be addressed through a dedicated compiler analysis phase that examines the Abstract Syntax Tree to identify uncaptured variables referenced within the lambda body.

\paragraph{Zero-Copy and Serialization-Free Data Transfer}
\toolname{} requires all data to be serialized for network transfer due to the inherent constraints of serverless function cloud interfaces.
Consequently, the framework makes no assumptions about data layout compatibility between the client's system and the function environment.
This problem is delegated to the serialization library to ensure portability.
Binary compatibility issues can arise in solutions that implement remote direct memory access and serialization-free data transfer~\cite{10.1145/3627703.3629568,copik2021rfaas}.
%
In such scenarios, the user bears responsibility for ensuring identical memory layouts between the client and function environments, which is a standard requirement for RDMA-accelerated and serialization-free data movement~\cite{273916}.

\section{Evaluation}

\label{sec:evaluation}

We demonstrate the ease of programming and parallel offloading in \toolname{} with micro-benchmarks,
thorough performance analysis using the Monte-Carlo approximation of a PI as a basis of discussion, and case studies using two applications: N-Queens from the Barcelona OpenMP Task Suite (BOTS)~\cite{duran2009barcelona} and a CPU Ray-Tracing application~\cite{RTW1}.
We used a \code{m5.4xlarge} virtual machine instance in the AWS \emph{eu-central-1} region, with 64 GiB RAM, network bandwidth of up to 10 Gb/s, and costing \$0.92 per hour.
The machine runs Ubuntu 22.04 and Intel Xeon Platinum 8175M CPU, with a frequency of 2.50GHz and 16 vCPUs, where each vCPU unit corresponds to a logical CPU core.
The virtual machine costs \$0.92 per hour.
We compile all benchmarks with the \toolname{} compiler based on the development branch of Clang 15.
When compiling, we use the CMake's \emph{Release} mode, which enables the \code{O3} optimization level in Clang.
For all benchmarks except serialization, we repeat measurements 21 times and reject the first warmup run to exclude the effects of initialization, cold startups, and connection setup.

In the evaluation, we answer the following questions:
\begin{itemize}
	\item How do the two underlying components of \toolname{} perform - serialization and invocation dispatch (Section~\ref{sec:micro-bench})?
	\item How does our performance compare against multi-threaded computations on the virtual machine  (Section~\ref{sec:eval-pi}-\ref{sec:eval-raytracing})?
	\item How do single-source functions in \toolname{} perform against a manual implementation
	      using cloud provider SDK (Section~\ref{sec:eval-pi-strong})?
	\item Can C++ benefit from fine-grained functions with independent configuration (Section~\ref{sec:eval-pi-mem})?
	\item Are cold startups a major performance issue in C++ functions (Section~\ref{sec:eval-pi-cold})?
	\item How large are the generated function binaries and deployment packages (Section~\ref{sec:eval-binary})?
\end{itemize}

\subsection{Microbenchmarks}\label{sec:micro-bench}
We use microbenchmarks to analyze the performance of two critical parts of \toolname{} runtime: data serialization with the \code{cereal} library and invoking functions with the internal HTTP client.
The choice of serialization format directly impacts the performance of serverless offloading,
and the function dispatch must scale up to thousands of invocations to allow for parallel processing.

\begin{figure}[t]
	\centering
	\begin{subfigure}{0.47\linewidth}
		\begin{tabular}{llccc}
			                        &        & Time  & Throughput \\ &&[ms] &[GiB/s]     \\\midrule
			\multirow{2}{*}{Binary} & Encode & 0.75  & 4.961      \\
			                        & Decode & 0.79  & 4.743      \\
			\multirow{2}{*}{\makecell{Binary                      \\Base64}} & Encode & 3.4 & 1.096  \\
			                        & Decode & 13    & 0.287      \\
			\multirow{2}{*}{JSON}   & Encode & 74.61 & 0.05       \\
			                        & Decode & 75.88 & 0.049      \\
		\end{tabular}
		\caption{Serialization of an array of 32-bit integers.}
		\label{fig:serialization-array-uint64}
	\end{subfigure}
	\begin{subfigure}{0.47\linewidth}
		\begin{tabular}{llccc}
			                                 &        & Time [ms] & Throughput \\ &&&[GiB/s]     \\\midrule
			\multirow{2}{*}{Binary}          & Encode & 21.65     & 1.076      \\
			                                 & Decode & 18.69     & 1.246      \\
			\multirow{2}{*}{\makecell{Binary                                   \\Base64}} & Encode & 34.89 & 0.667  \\
			                                 & Decode & 46.51     & 0.501      \\
			\multirow{2}{*}{\makecell{JSON}} & Encode & 278.45    & 0.084      \\
			                                 & Decode & 289.07    & 0.081      \\
		\end{tabular}
		\caption{Serialization of an array of structures.}
		\label{fig:serialization-array-struct}
	\end{subfigure}
	\caption{Benchmarking the performance of data serialization in \toolname{} with \code{Cereal} library. REST-based serverless platforms can only accept payload as a JSON or base64-encoded binary.}
	\label{fig:serialization}
\end{figure}

\paragraph{Serialization}\label{sec:micro-bench-serialization}
Serverless systems put constraints on the type of data that can be transmitted to a function.
In the case of AWS Lambda, the integration with API gateway restricts the types.
In practice, functions can be invoked with a valid JSON object or Base64-encoded binary data.
We evaluate two scenarios of serializing C++ objects with the \code{cereal} library:
the default one using a direct JSON serialization, and an alternative one that performs a binary
serialization followed by the Base64 encoding.
As a baseline, we use the plain binary serialization that can be used in systems using RPC
and RDMA requests~\cite{copik2021rfaas}.
On all benchmarks, we exclude the cost of memory allocation for the resulting payload (serialization)
and data objects (deserialization).
To create a realistic experiment environment, we exclude the influence of CPU caches and measure
the performance of these operations when reading data from main memory.
To that end, we use the Intel intrinsic \code{\_mm\_clflush} to flush and invalidate cache lines storing serialization objects.
We repeat measurements 1,001 times and reject the first warmup execution.

First, we benchmark the serialization of an array of one million 32-bit unsigned integers (Figure~\ref{fig:serialization-array-uint64}), with a total binary size of 4 MB (3.81 MiB)
which is close to the maximum input payload of 6 MB on AWS Lambda.
Adding the Base64 encoding decreases the serialization throughput by 4.52x.
However, the binary option still significantly improves performance against a plain JSON serialization that has to convert numbers into their string representation.
Then, we evaluate the serialization of a structure consisting of two integers and a single string with 42 characters, with a custom serialization method provided to the library (Figure~\ref{fig:serialization-array-struct}).
This serialization is more expensive due to pointer jumping and more complex encoding of a \code{std::string}.
Nevertheless, our custom binary serialization format is up to 8x faster than a vanilla JSON serialization.

\begin{figure}[t]
	\centering
	\includegraphics[width=\textwidth]{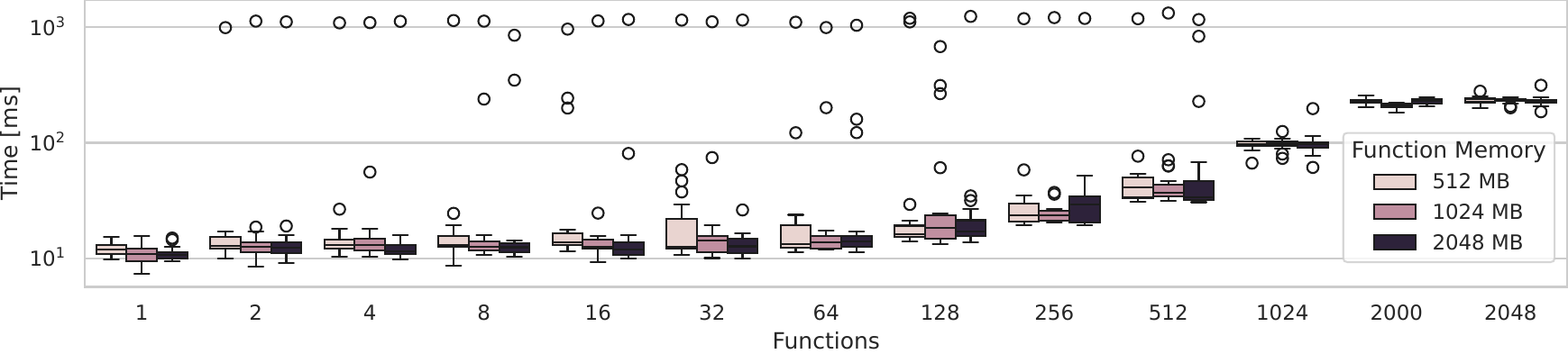}
	\caption{The total latency of concurrent invocations of warm AWS Lambda functions with the nghttp2 dispatcher, as observed by the client. Boxes represent the data between the first and third quartiles, with whiskers adding 1.5 of the interquartile range.}
	\label{fig:aws_http_client_latency}
\end{figure}

\paragraph{AWS Lambda Client}
We examined the latency of parallel invocations to AWS Lambda functions when using our HTTP2 dispatcher in \toolname{}.
Our dispatcher implementation uses Boost.Asio abstractions to dispatch dozens of parallel invocations without allocating a separate thread for each task.
We use the default configuration with 16 HTTP/2 connections and invoke a no-op C++ function deployed with our compiler.
We use an AWS cloud account with a maximum number of concurrent function invocations of 2,000.

Figure~\ref{fig:aws_http_client_latency} shows that the median latency of a single invocation is 12 and 11 milliseconds, depending on the function memory size.
All invocations are warm, and the median function execution time varies between 700 and 925 microseconds.
We do not observe a significant influence of the function memory size on parallel invocations.
Since the dispatcher can reuse the TCP connection, the initial latency of establishing an HTTPS connection to the AWS REST API is paid only in the first cold invocation.
After the connection initialization, the client dispatches invocations until it reaches the AWS Lambda concurrency limit or exhausts client dispatcher resources.

\emph{Slowdown.}
The total client runtime increases at larger scales, from ca. 14 milliseconds at 64 invocations up to 230 milliseconds at 2000 invocations.
To verify if this slowdown is caused by the serverless system or resource limitations of the client,
we replicated the virtual machine 8 times to run one-eighth of the original workload from each machine, obtaining in total the same number of concurrent invocations.
We implemented an MPI-based dispatcher with a barrier to synchronize distributed clients, and measure the longest runtime across all ranks.
At 512 invocations (64 per MPI rank), the median runtime decreased from 34-41 ms to 28.5-38 ms, indicating that
runtime is slightly influenced by the limitations of the dispatcher or virtual machine resources.
%
%
However, at 1024 and 2000 invocations, the runtime was still up to 6.4x and 7.7x slower when executed by eight VMs concurrently instead of one,
thus indicating scaling limitations on the AWS Lambda system.

\emph{Boost.Beast.}
We also compared against the second dispatch implementation using Boost.Beast and independent HTTP requests.
We increased the system limit on opened file descriptors to support 2,048 concurrent invocations.
There, the lowest observed median latency was 13.45 milliseconds for a single function invocation.
However, while the client latency stays relatively constant for the \code{nghttp2} dispatcher until 256 and 512 invocations, it quickly increases for the \code{Beast} dispatcher, reaching 135 milliseconds on 64 invocations and over four seconds on 2,000 concurrent invocations.
Thus, the dispatcher provided with \toolname{} can offer an optimized implementation of remote functions without relying on the user to select the best-performing library and configuration.

\begin{figure}[t]
	\centering
	\includegraphics[width=\textwidth]{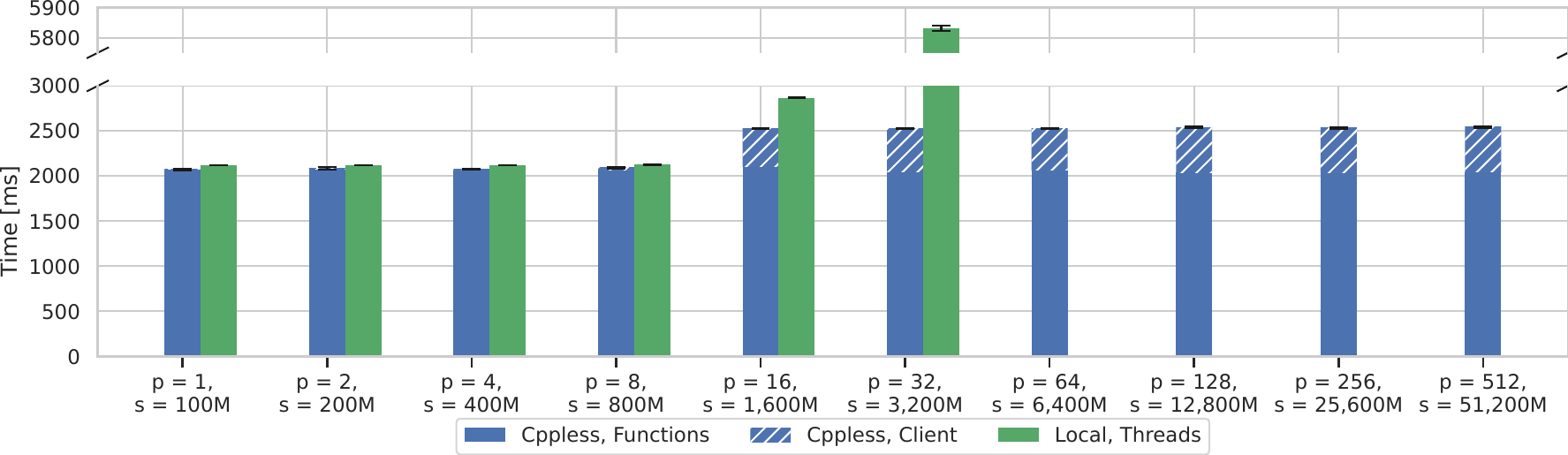}
	\caption{Weak scaling of PI approximation on AWS Lambda and the virtual machine. Bars represent the mean with 95\% confidence intervals.}
	\label{fig:pi_weak_scaling}
\end{figure}

\subsection{Case Study: Monte Carlo Pi Benchmark}
\label{sec:eval-pi}
To evaluate the scalability and elasticity of serverless computing with \toolname{},
we conducted four experiments with a perfectly parallelizable application: approximating the Pi number with Monte-Carlo simulation, as introduced in Section~\ref{sec:compilation}.
Unless specified otherwise, benchmarks are compiled with the addition of the \code{fast-math} flag
and functions use 2048 megabytes of memory.

\subsubsection{Weak Scaling}
First, we examine the weak scalability by increasing the number of parallel functions
up to 512 (Figure~\ref{fig:pi_weak_scaling}).
We compare the performance against computation on the virtual machine, using manual dispatch of results to threads.
Since the virtual machine is limited to the local 16 virtual CPUs, it cannot handle increasing workloads.
Instead, users can extend computing resources by dynamically dispatching tasks with \toolname{}, incurring only a small, constant overhead on the HTTP client.
The computing cost stays relatively constant: it changes by less than 3\% between problem sizes.
At 512 functions, the total cost is 500.4x larger than using only one function.

\begin{figure}[t]
	\centering
	\includegraphics[width=\textwidth]{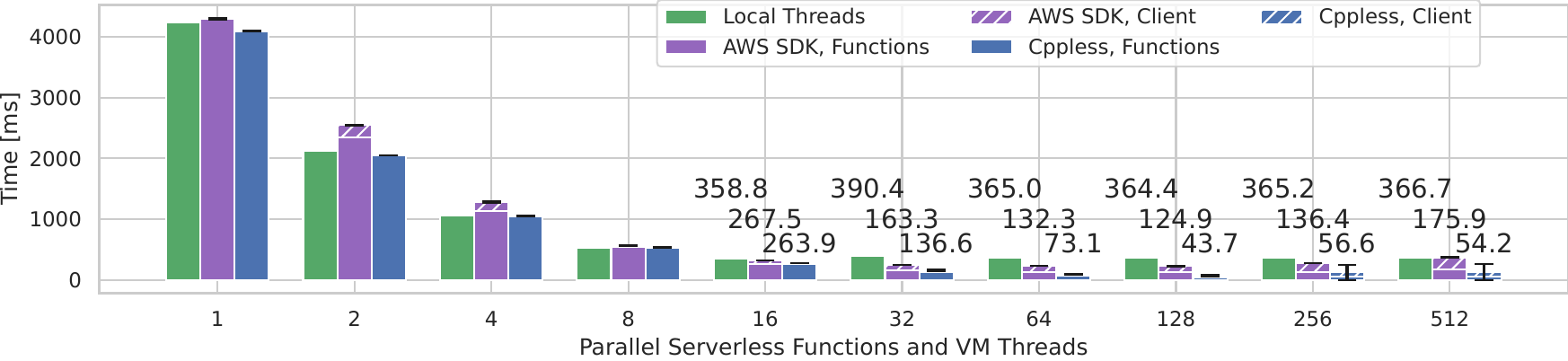}
	\caption{Strong scaling of PI approximation on AWS Lambda and the virtual machine, with 200 million samples. Bars represent the mean with 95\% confidence intervals.}
	\label{fig:pi_strong_scaling}
\end{figure}
\begin{figure}[t]
	\centering
	\includegraphics[width=\textwidth]{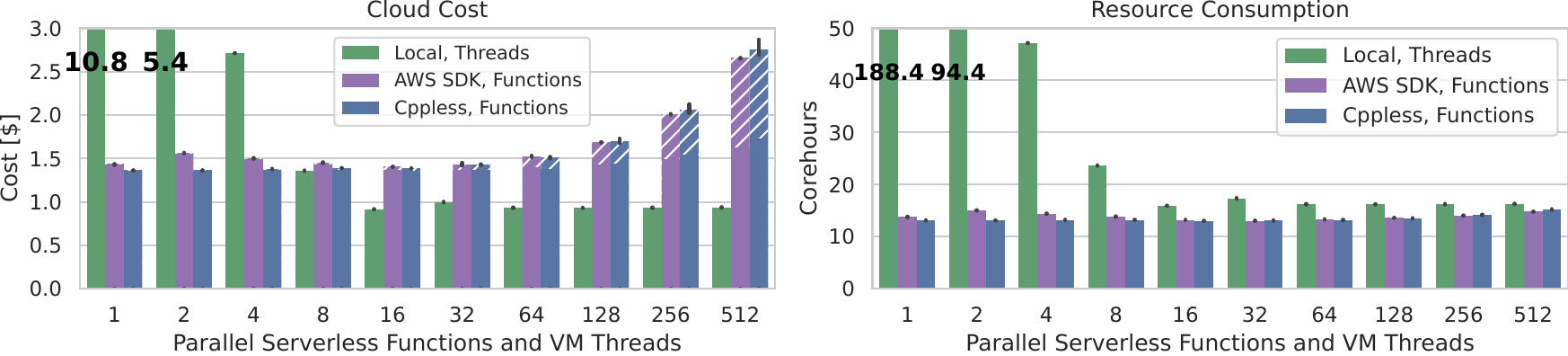}
	\caption{Cost of 10,000 repetitions of PI benchmark (Figure~\ref{fig:pi_strong_scaling}); computing cost (solid) and invocation fee (dashed), mean with 95\% confidence interval.}
	\label{fig:pi_strong_scaling_cost}
\end{figure}

\subsubsection{Strong Scaling}
\label{sec:eval-pi-strong}
We examine the efficiency by scaling the Pi computation with 200 million samples.
We compare against two baselines: multithreaded scaling on the virtual machine,
and a manual serverless implementation using the official AWS SDK. 
By comparing against the latter, we can verify that \toolname{} can extract and deploy C++ functions without introducing overheads.

\emph{Performance.}
Dispatching computations with \toolname{} scales efficiently until 128 functions, where the median runtime of a function decreases to 42.7 milliseconds.
There, the overheads of function scheduling and network protocols impact the further scalability (Figure~\ref{fig:pi_strong_scaling}).
The manual baseline scales worse, which can be explained by the underlying implementation that uses a fixed number of HTTP connections and creates a new thread for each asynchronous invocation.\footnote{The issue has been confirmed by the AWS team: \url{https://github.com/aws/aws-sdk-cpp/issues/2991}; accessed on 24.06.2024.}
We experimented by increasing the fixed number of TCP connections used by the SDK, but
it did not improve the performance and significantly decreased stability.

\emph{Cost.}
To compare the costs, we make the optimistic assumption that the virtual machine is charged only for the time spent on the Monte-Carlo computation.
%
In serverless, the cost is split into two components: function cost for memory and CPU time, and a flat fee for each invocation.
The results show that offloading with \toolname{} enables scaling beyond VM limits with a slow increase in cost (Figure~\ref{fig:pi_strong_scaling_cost}).
The increase is caused by the invocation fee charged for each parallel function, which impacts computations with very short and fine-grained functions; at 512 workers, the median duration of a single function is just 28.4 milliseconds.
At full utilization, local computations are cheaper because the bare price of a vCPU on a virtual machine is lower than that of Lambda.\footnote{In the \code{m5.4xlarge} virtual machine, a single vCPU costs \$0.0575 per hour. AWS Lambda has an equivalent of 1 vCPU at 1,769 MB of memory, which would cost \$0.104 per hour - 1.8x more.}

However, functions consume fewer core hours than a VM computation.
Serverless can perform better when the cloud provider supplies a better CPU.
Furthermore, we notice that the variance of serverless cost increases with the number of workers,
while resource consumption is stable.
The larger variance is explained by an increased frequency of cold starts (up to 2.5\% of all invocations across repetitions).
At 512 functions, a cold invocation takes 9.8 ms, which is 7\% longer than a warm one, but adds ca. 13 ms of initialization time.
Since AWS Lambda charges for the initialization time, a cold invocation at this scale costs up to 2.4x more than a warm one.
When spurious cold starts occur, the total cost is up to 18\% higher, but these only happen in 3.5\% of all repetitions.

\begin{figure}[t]
	\centering
	\begin{subfigure}{0.45\linewidth}
		\includegraphics[width=\textwidth]{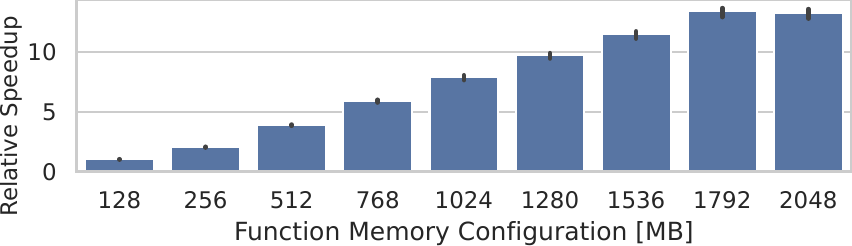}
		\caption{Function runtime speedup.}
		\label{fig:pi_memory_scalability_performance}
	\end{subfigure}
	\hfill
	\begin{subfigure}{0.45\linewidth}
		\includegraphics[width=\textwidth]{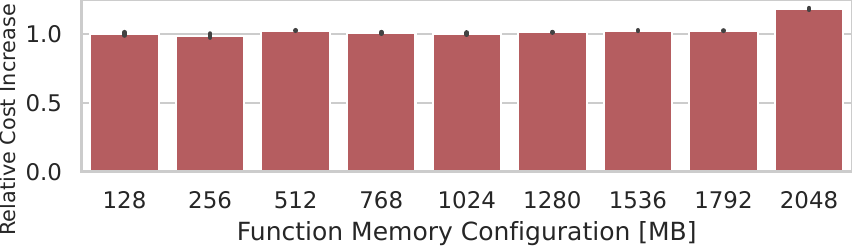}
		\caption{Function cost increase.}
		\label{fig:pi_memory_scalability_cost}
	\end{subfigure}
	\caption{Performance of the PI benchmark with a varying resource allocation for the serverless function. Speedup and costs are normalized to results of 128 MB function.}
	\label{fig:pi_memory_scalability}
\end{figure}

\subsubsection{Memory Scalability}
\label{sec:eval-pi-mem}
Serverless frameworks must choose between deploying many distinct Lambda functions or fusing them together.
The latter simplifies dynamic code dispatch in systems like Lithops and Crucial and decreases the frequency of cold starts, as each generic function instance can handle all tasks.
However, having only a single cloud function type decreases flexibility, as resource allocation for the function has to be decided upfront.

\toolname{} deploys each offloaded C++ function as a separate Lambda function, allowing to configure memory and CPU resources independently for each task.
To verify the gains of this policy, we scaled the memory allocation of the Pi function
(Figure~\ref{fig:pi_memory_scalability}).
In this compute-intensive task, scaling up to a full logical core provides faster computation for the same price.
On the other hand, prior works show that other serverless functions achieve the best efficiency at much lower memory allocations~\cite{copik2021sebs}.
With a single cloud function type, users would waste resources on memory-bound functions or lose the opportunity to accelerate compute-intensive workloads for free.
Thus, deploying many function types in \toolname{} allows efficiently mixing different operations types in a single application.

\begin{figure}[t]
	\adjustbox{max width=\textwidth}{
		\centering
		\begin{tabular}{lccccccccccc}

			\multirow{2}{*}{\makecell{Function                  \\Memory}}
			 & \multirow{2}{*}{C++}
			 & \multicolumn{2}{c}{Python 3.8}
			 & \multicolumn{2}{c}{Python 3.9}
			 & \multicolumn{2}{c}{Python 3.10}
			 & \multicolumn{2}{c}{Python 3.11}
			 & \multicolumn{2}{c}{Python 3.12}                  \\

			\cmidrule(lr){3-4}
			\cmidrule(lr){5-6}
			\cmidrule(lr){7-8}
			\cmidrule(lr){9-10}
			\cmidrule(lr){11-12}

			 &                                 & Bare   & NumPy
			 & Bare                            & NumPy
			 & Bare                            & NumPy
			 & Bare                            & NumPy
			 & Bare                            & NumPy          \\\midrule

			128
			 & 10.4
			 & 123.12                          & 607.6
			 & 74.06                           & 498.08
			 & 73.31                           & 311.9
			 & 73.52                           & 490.79
			 & 81.1                            & 526.84         \\

			512
			 & 10.36
			 & 122.5                           & 621.08
			 & 74.4                            & 489.97
			 & 71.43                           & 311.76
			 & 74.29                           & 498.9
			 & 78.73                           & 527.48         \\

			1024
			 & 10.82
			 & 124.61                          & 614.76
			 & 73.21                           & 496.04
			 & 73.31                           & 310.33
			 & 72.53                           & 495.11
			 & 79                              & 533.87         \\

			2048
			 & 11.78
			 & 124.34                          & 594.1
			 & 72.58                           & 471.84
			 & 72                              & 297.59
			 & 73.94                           & 473.15
			 & 79.24                           & 513.24         \\
		\end{tabular}
	}
	\caption{
		The sandbox initialization time on cold startup (milliseconds) at different memory configurations (megabytes). We compare the Cppless Pi function against Python functions that return the input payload.}
	\label{fig:cold-startup}
\end{figure}

\subsubsection{Cold Startups}
\label{sec:eval-pi-cold}
Serverless is designed for irregular and elastic workloads, and launching many parallel functions leads to cold startups, as each function instance needs its own container.
Since \toolname{} deploys functions separately for higher resource efficiency (Section~\ref{sec:eval-pi-mem}), it has to tolerate more frequent cold invocations.
Thus, we examine the performance effects of a cold startup on C++ functions.
We compare against functions implemented in Python, the most popular language in serverless computing (Section~\ref{sec:background:lang}).
We implemented a simple Python function that returns the payload and its cold status.
However, pure Python cannot provide performance that would make it competitive in this Monte-Carlo simulation, requiring libraries such as NumPy.
Thus, we use a second variant of this function that only adds a single NumPy import.

We triggered cold initialization by changing the function's environment configuration, and gathered the cost of initializing the function sandbox as reported by the cloud provider (Figure~\ref{fig:cold-startup}).
Thanks to the compilation and a lack of interpreter sessions, C++ functions can be initialized
very quickly, with only 10-12 milliseconds of added overhead.
While the cost of initializing a bare Python function can be up to 11.8x higher than in C++,
adding a single popular library needed for numerical computations increases the overhead to 25.26 - 58.4x.
Thus, \toolname{} helps users to easily integrate serverless functions into C++ applications, where the runtime overhead of using many fine-grained and specialized functions is minimal.

\subsubsection{Code Changes}
In \toolname{}, users need to add two lines of configuration and modify 7 lines of code
to dispatch tasks and wait for the results.
When implementing the same benchmark with the official cloud SDK, users need to manually serialize data and implement callbacks for asynchronous invocations.
There, developers need at least 4 lines of code for configuration, 16 lines for the dispatch loop, and 11 lines for the callback handling results.
On top of that, users need to separately build, package, and deploy the function and later hardcode the function name in the main application.

\begin{figure}[t]
	\centering
	\includegraphics[width=\textwidth]{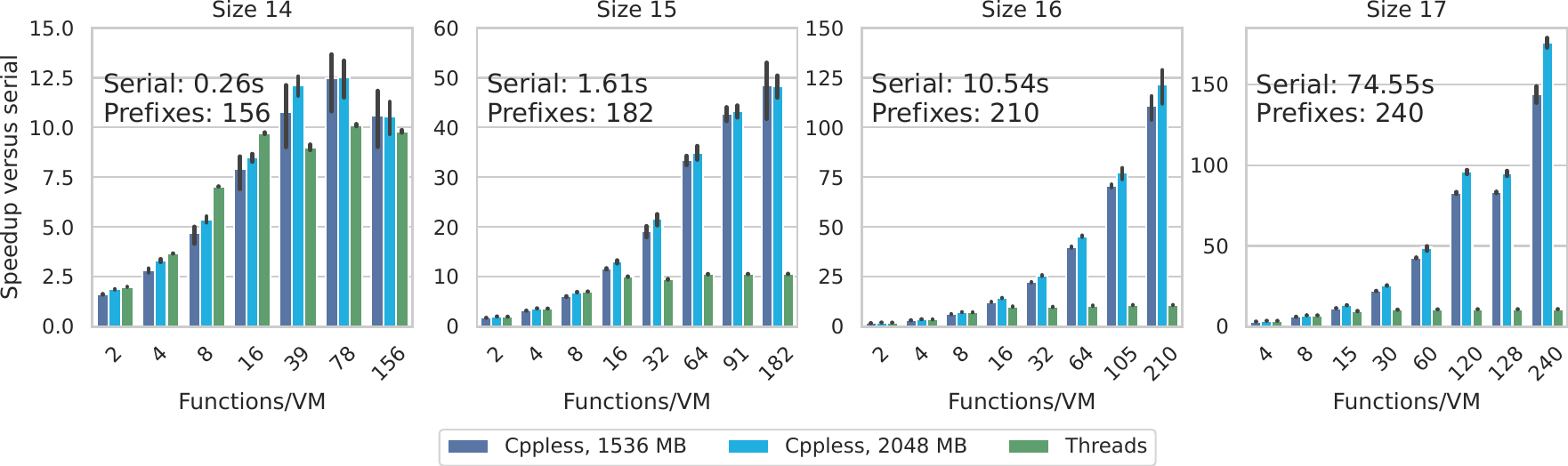}
	\caption{N-Queens benchmark with varying board size. We measure the entire processing time, including prefix computation. Mean speedup against the serial runtime, with 95\% confidence intervals.}
	\label{fig:nqueens-perf}
\end{figure}

\subsection{Case Study: OpenMP Benchmark}
\label{sec:eval-openmp}
Next, we evaluate \toolname{} with the  N-Queens problem from the BOTS suite.
In this benchmark, we find a placement of \emph{N} queen figures on a chessboard of size \emph{NxN}, such that no two queens threaten each other.
The problem is known to be NP-hard and has numerous solutions dependent on the size of the board \emph{N}.
%
%
We changed the array-based implementation from the BOTS benchmark suite with one that uses bit patterns to represent board states~\cite{richards1997backtracking}, improving the performance of determining queen placement.
We parallelize the solution by pre-computing \emph{prefix} tasks of length $p$ where the location of the first $p$ queens is fixed, allowing us to decompose the primary problem into smaller tasks~\cite{kise2004solving}.
The local parallel implementation requires 35 lines of code, while a solution with the serverless \toolname{} dispatcher is implemented with 47 lines of code.
No additional serialization is involved, as prefixes are sent using the \code{std::vector} class.

\begin{figure}[t]
	\includegraphics[width=\textwidth]{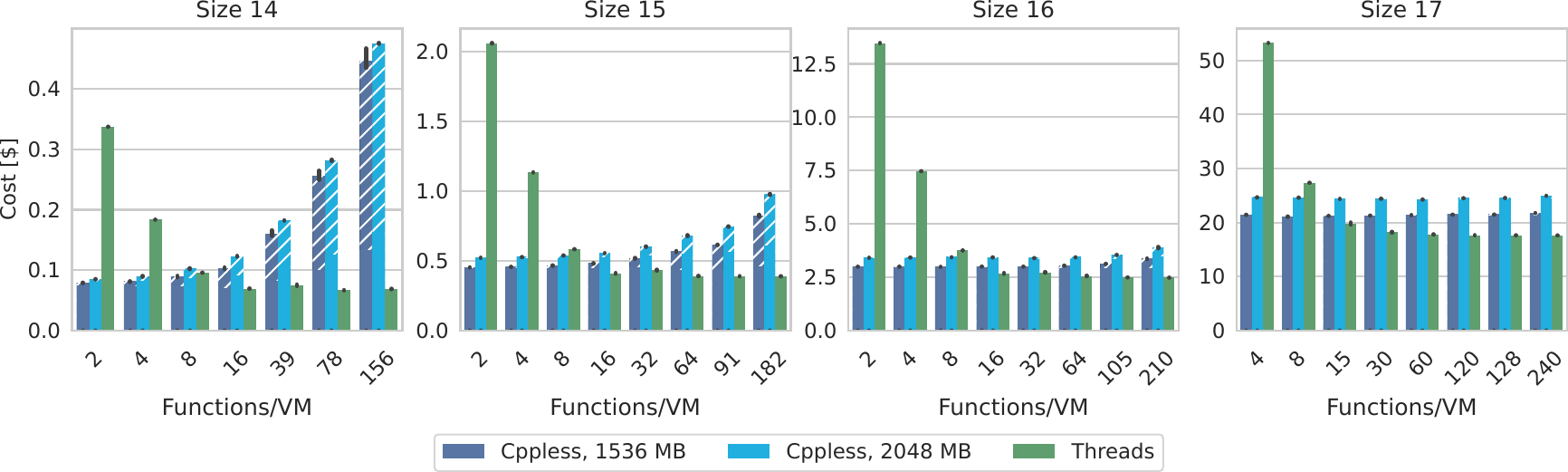}
	\caption{Cost of 10,000 repetitions of N-Queens benchmark (Figure~\ref{fig:nqueens-perf}); mean with 95\% confidence interval. \toolname{} includes computing cost (solid), invocation fee (dashed), and dispatch cost of host vCPU (black top bar), which does not exceed 0.2\% of total cost.}
	\label{fig:nqueens-cost}
\end{figure}

\emph{Performance.}
We compare the local computation with the dispatch to serverless when using a prefix of length 2 (Figure~\ref{fig:nqueens-perf}).
At the largest scale, each function or thread receives a single prefix to resolve.
Offloading computations to serverless functions provides speedups of up to 111x and 122x for $N=16$, and up to 144x and 176x for $N=17$.
However, the results indicate that serverless offloading does not achieve linear scaling.
This can be explained by the variance in the workload assigned to different tasks, with the total execution time limited by the longest-running task.
On a large scale, the slowest task takes 2.38-5.3x longer than the fastest one ($N=16, 17$) and even up to 151x longer on smaller problem sizes.
Furthermore, at smaller problem sizes and larger scales, dispatching tasks can take up to 14\% of the total time.
Nonetheless, the users benefit from the pay-as-you-go billing mode in such a heterogeneous workload: parallel functions do not wait for the results of longer-running functions and thus do not accrue charges.

\emph{Cost.}
We compare the total computation cost of local and serverless parallelization.
As before, the cost of serverless functions is split into two components: the function cost obtained by querying the cloud billing data and the invocation fee.
We add a third cost component to obtain a fair comparison against local multithreading: host time, which includes the additional time needed for data preparation and task serialization.
We define it as a time from the beginning until all functions are dispatched, and we multiply it by the cost of using a single vCPU of the virtual machine.

As seen previously, a fully utilized virtual machine delivers lower cost due to an increased compute price at AWS Lambda (Figure~\ref{fig:nqueens-cost}).
More importantly, we notice that the total cost at the two largest scales increases only by up to 14\% and 2\% for $N=16, 17$, respectively.
Thus, as in the previous benchmark, users can obtain significant speedup at a minor cost increase.

\emph{Knapsack and Floorplan}
We also tested \toolname{} with two other benchmarks from the BOTS suite.
However, these rely on recursive task creation and use shared memory to synchronize results between tasks to prune early branches.
Since available serverless systems support neither distributed memory nor inter-function communication, these benchmarks cannot scale efficiently there.
Thus, we focus instead on workloads that fit better this programming model.

\subsection{Case Study: Ray-Tracing}
\label{sec:eval-raytracing}
As a second application, we consider a Monte-Carlo implementation of ray tracing~\cite{RTW1} with a bounding volume hierarchy mechanism~\cite{RTW2}.
The implementation consists of nine translation units with 1,616 lines of C/C++ code,
and the offloading includes dynamic lists of objects, custom and polymorphic types, and shared pointers.

The benchmark scenario renders a random scene divided into smaller tiles to create parallel tasks.
In local multi-threading, worker threads can access image tiles directly in the memory.
For \toolname{}, we compute the bounding volume hierarchy locally and the serverless
function is invoked with the image tile and volume hierarchy.
For each benchmark, we measure the entire ray-tracing process, including splitting images into tiles and serialization, and exclude the time needed for the random generation of the initial image.

\begin{figure}[t]
	\centering
	\includegraphics[width=\textwidth]{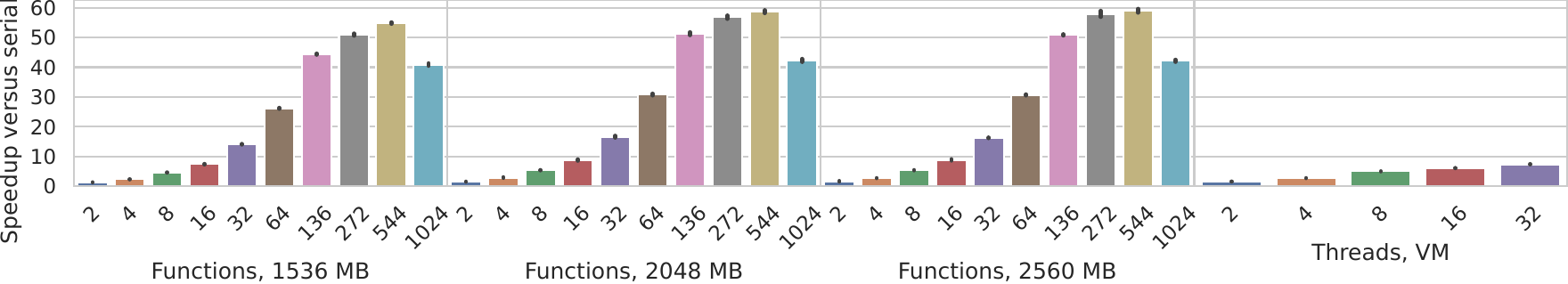}
	\caption{CPU-Raytracer rendering of a 600x600 image. We measure the entire runtime, including image partitioning. Mean speedup of 20 repetitions against the serial runtime with one thread of 59.159 seconds.}
	\label{fig:ray_500}
\end{figure}

\paragraph{Performance}
We evaluate the benchmark on an image of size 600x600, varying the tile dimensions to control the number of dispatched functions.
The image tile was changed from 600x300 for two functions to 19x19 for 1024 functions.
\toolname{} provides a speedup of up to 59x, while the limited resources of a non-elastic virtual machine saturate quickly (Figure~\ref{fig:ray_500}).
This result is already attractive for the end user, as their compute-intensive task taking almost 60 seconds can now be completed in slightly more than one second without allocating a dedicated and more powerful virtual machine with multiple cores.

However, the obtained speedup is lower than the number of invoked functions.
First, we must create the bounding volume hierarchy, a serial process on the host that adds, on average, 4.2 milliseconds to the runtime and increases payload size.
However, the main culprit of lower scalability is the unequal work distribution.
This uneven workload distribution is caused by the unequal division of the image to selected tile sizes and the varying per-pixel workloads, as the computation time of each task varies and depends on the objects present in the assigned tile.
For example, on 136 tiles, the longest-running task can take 27.8x more time than the fastest.
There, the contribution of function duration to the total computation time can vary between 3.7\% and 94.8\%.
%
Finally, the speedup decreases when the number of functions increases from 544 to 1024.
This can be explained by the time needed to serialize and dispatch functions, which grows from 0.52 to 0.99 seconds, while the maximum function duration decreases from 402 to 228 ms (2560 MB memory).

\begin{figure}[t]
	\includegraphics[width=\textwidth]{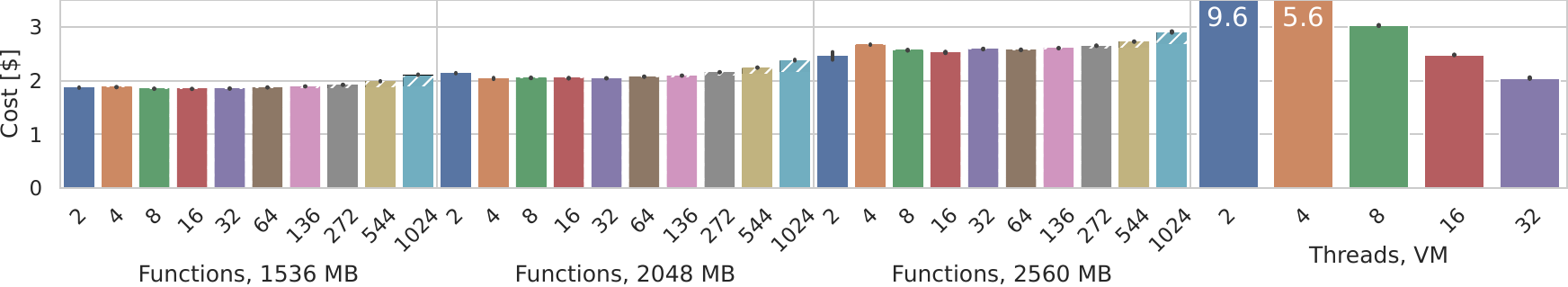}
	\caption{The total cost of 1,000 invocations of the CPU-Raytracer benchmark from Figure~\ref{fig:ray_500}. For serverless, we include the computation cost (solid), invocation fee (dashed), and cost of host preparation (black top bar) which does not exceed 0.8\% of total cost.}
	\label{fig:ray-cost}
\end{figure}

\paragraph{Cost}
With \toolname{} functions, we significantly decrease processing time with only a minor increase in the total cost (Figure~\ref{fig:ray-cost}).
%
Particularly in the case of functions configured with 1536 MB of memory, 544 functions decrease
computing time by up to 54.8x.
However, the total cost increases from 2 to 544 functions by only 6\%.
Thanks to our embedding of elastic cloud functions into C++ applications, users effectively achieve faster processing almost for free.

\paragraph{Code Modifications}
To understand the difference between \toolname{} and OpenMP, we compared our interface with a ray tracing application based on OpenMP offloading~\cite{raytracerompcloud,mortattiAutomaticRayTracerCloud2018}.
We replace parallel directives with an explicit dispatch interface and move the loop body into a C++ lambda function.
However, the main difference lies in data transmission that uses serialization instead of \texttt{target map} directives.
Rather than specifying OpenMP \texttt{declare mapper} or using plain arrays, we implement twelve serialization routines to support custom datatypes.
Of these, eleven simply enumerate serialized class fields.
As a benefit, we can transmit C++ vectors and shared pointers without additional modifications, since all variables captured by the lambda are implicitly serialized.

\subsection{Code Size}
\label{sec:eval-binary}

\begin{figure}[t]
	\adjustbox{max width=\textwidth}{
		\centering
		\begin{tabular}{lccccccccccc}

			\multirow{2}{*}{Benchmark} & \multicolumn{3}{c}{Function} & \multirow{2}{*}{\makecell{Dynamic                                                                       \\Dependencies}} & \multirow{2}{*}{\makecell{Function\\Data}} & \multicolumn{2}{c}{Zip Package} \\

			\cmidrule(lr){2-4}
			\cmidrule(lr){7-8}

			                           & Object Files                 & Static Links                      & Final Entrypoint &           &          & Compressed & Uncompressed \\
			\midrule
			NQueens                    & 920 kB                       & 15.08 MB                          & 5.89 MB          & 0         & 0        & 8.33 MB    & 19.8 MB      \\
			Ray Tracing                & 4.3 MB                       & $''$                              & 6.45 MB          & 0         & 0        & 8.44 MB    & 20.19 MB     \\
			SeBS Thumbnailer           & 594 kB                       & $''$                              & 5.77 MB          & 49.37 MB  & 0        & 25.5 MB    & 69.18 MB     \\
			SeBS Image Recognition     & 642 kB                       & 17.33 MB                          & 5.8 MB           & 479.16 MB & 97.96 MB & 215.1 MB   & 595.38 MB    \\
			SeBS Graph Pagerank        & 588 kB                       & 24.5 MB                           & 6.68 MB          & 0         & 0        & 20.32 MB   & 52.73 MB     \\
		\end{tabular}
	}
	\caption{
		Binary size of serverless functions: object files are linked with static dependencies to produce the final function. Functions are deployed as a zip package, together with dynamically linked libraries and additional function data.
		The image recognition cannot be deployed as a package on AWS Lambda due to its size.
	}
	\label{tab:code-sizes}
\end{figure}

Since \toolname{} does not apply dedicated solutions for selective compilation (Section~\ref{sec:implementation_limitations}), we verified if the generated binary size adds significant overhead to deployment.
For each application, we analyze the sizes of generated alternative entry points and final deployment packages uploaded to the cloud.
In addition to the NQueens benchmark and Ray Tracing application, we used C++ ports of three serverless functions from the SeBS benchmark suite~\cite{copik2021sebs}: image thumbnailer with OpenCV 4.5; image recognition with ResNet-50, OpenCV, Torch 1.1, and Torchvision 0.1; and graph processing with igraph 0.10.
We adapt each function for offloading with \toolname{} and generated deployment packages.
Figure~\ref{tab:code-sizes} shows that dynamic libraries dominate the size of serverless deployment.
These include libc and its dependencies, which the AWS Lambda packaging script includes by default for compatibility.
Alternative entry points are statically linked with Lambda runtime, libcurl, and \toolname{} dependencies like nghttp2.
In this case, pruning unnecessary symbols used only on the host side can help reduce the size of the final entry point.
However, deploying practical serverless applications is likely to be dominated by complex dependencies like OpenCV and Torch.
In particular, the image recognition benchmark can only be deployed as a Docker container due to the large size of the Torch library.

\section{Related Work}
\label{sec:related_work}
Prior approaches focus on solutions that exploit dynamic and interpreted languages (Section~\ref{sec:related_interpreted}),
solutions that extend the OpenMP programming model to support HPC clusters and clouds (Section~\ref{sec:related_omp}),
and offload C++ functions to accelerator devices (Section~\ref{sec:related_gpu}).
\toolname{} takes a different approach to support distributed deployment in the language
and does not rely on additional runtimes like MPI or Spark,
and general-purpose cloud offloading (Section~\ref{sec:related_cloud_offload}).

\subsection{Interpreted Languages}
\label{sec:related_interpreted}
Many prior systems focused on interpreted and high-level languages, which implicitly support serializing data objects, allow for dynamic code extraction, and provide reflection support for runtime dispatch.
Kovachev et al.~\cite{kovachevFrameworkComputationOffloading2012} proposed offloading compute-intensive Android services to a Java server to speed up computations and reduce energy usage.
There, Java code is shipped at the runtime to a persistent server in the cloud and cached for reuse in future executions.
Lithops~\cite{9218932} is a multi-cloud framework for offloading Python functions to serverless functions.
It implements a replacement of the standard \code{multiprocessing} library,
which provides the same interface
but invokes remote cloud functions instead of spawning a new local OS process.
The function code is not deployed to the cloud ahead of time; Lithops analyzes it at runtime to detect all dependent modules, serializes them, and sends them to the cloud storage.
The code is later fetched and executed by a generic serverless Python worker.
Python functions can use modules and libraries embedded in the default container image~\cite{lithopsDocs}.
To exploit additional libraries, users must manually build and deploy extended container images before using the software.

Crucial implements stateful serverless applications in Java~\cite{10.1145/3361525.3361535}.
Functions invocations are represented as remote threads that implement the standard \code{Runnable} interface,
providing an abstraction layer similar to the standard interface of Java threads.
Functions are invoked with the payload specifying the name of the class implementing the \code{Runnable} interface and its parameters.
The serverless executor can create an instance of the class thanks to the availability of reflection in Java, which is missing in C++.
Crucial deploys a single Lambda function per application~\cite{crucialExamples}, preventing fine-grained memory and CPU resource configuration across different Java functions.
Crucial implements a custom storage system presented to users as a distributed shared object layer, which allows them to manage function state and receive
data from offloaded tasks since the \code{Runnable} interface does not support the direct return of results.
Thus, it cannot be deployed to vanilla serverless platforms.
Serverless Shell~\cite{10.1145/3491084.3491426} builds on top of Crucial's distributed shared object layers,
executing shell scripts remotely on serverless functions.
However, the code intended for remote execution must be explicitly invoked through a dedicated executor.

Kappa~\cite{10.1145/3419111.3421277} targets long-running serverless applications and implements automatic checkpointing for Python functions.
A dedicated compiler transforms user code by inserting continuations,
but its support is limited to native Python code and does not handle Python C extensions.
Containerless~\cite{herbert2020languagebased} compiles a subset of JavaScript into Rust, providing speculative and opportunistic acceleration of functions implemented in
high-level languages.

\subsection{OpenMP}
\label{sec:related_omp}
OpenMP was initially designed for multithreaded computations in shared-memory systems.
Multiple attempts have been made to extend OpenMP with offloading computations to a distributed and remote system, and these can be classified as \emph{cluster} and \emph{cloud} OpenMP.
Later iterations of the standard added support for offloading computations to a \emph{target} device, with a particular focus on accelerators (Section~\ref{sec:related_gpu}).

\paragraph{Cluster OpenMP}

In this mode, data and computations are distributed among nodes of an HPC cluster.
These attempts were often limited by the overhead of sharing memory across computing nodes.
Sato et al. proposed Omni OpenMP~\cite{satoDesignOpenMPCompiler},
where the compiler inserts communication routines into the code to guarantee the correctness and consistency of shared data.
In 2006, Intel offered Cluster OpenMP that used memory page protection
to detect cross-node memory accesses and enforce synchronization.
However, initial evaluation by Terboven et al. demonstrated scalability issues, with basic OpenMP primitives slower by up to four orders of magnitude~\cite{terbovenFirstExperiencesIntel2008}.
Cluster OpenMP has been discontinued in 2010~\cite{clusteropenmp}.
XcalableMP (XMP) is a programming model for clusters that uses compiler directives similar to OpenMP, with data represented as distributed arrays in the Partitioned Global Address Space (PGAS) model~\cite{nakaoProductivityPerformanceGlobalView2012}.
XcalableACC~\cite{nakaoXcalableACCExtensionXcalableMP2014} extended later XMP with OpenACC directives to support accelerator programming, focusing on clusters with HPC interconnects like InfiniBand and Tightly Coupled Accelerators (TCA).
This solution relies on a custom Omni compiler that translates XMP directives into functions to a runtime library.

Several works attempted to combine OpenMP and distributed computations with MPI by exploiting OpenMP 4.0 offload.
Jacob et al.~\cite{jacobExploitingFineCoarseGrained2015} was the first to use OpenMP offload to target remote CPUs.
Instead of defining a global address space, cluster nodes are treated as a collection of disjoint shared-memory address spaces.
The implementation is based on LLVM, with kernel code included in a fat binary with host code.
The actual runtime is built on top of MPI, with the host communicating with worker nodes to distribute computation tasks.
Recently, Yviquel et al. proposed OpenMP Cluster (OMPC)~\cite{yviquelOpenMPClusterProgramming2022}, a tasking model for HPC clusters.
They implement a new device plugin for LLVM's OpenMP,
with a dedicated management module for automatic data movement and handling task dependencies across nodes.
The host and target code are also compiled into a fat binary, and communication is done with MPI.
Keftakis et al.~\cite{keftakisExperiencesTaskbasedProgramming2022} implemented offloading where each cluster node becomes a separate OpenMP device.
Their approach is based on the OMPi compiler and generates a single executable that contains the host code and all target kernels.
Remote communication is delegated to MPI by spawning new MPI processes as copies of the host process, which avoids the problem of dispatching code at runtime.

Cluster OpenMP solutions tend to assume a homogenous system where a single copy of an application executes on different nodes.
In contrast, \toolname{} has to generate multiple targets that are separately uploaded to a cloud platform.
Instead of modeling external resources as devices that often need to be predefined ahead of time through a configuration file~\cite{keftakisExperiencesTaskbasedProgramming2022}, we dynamically determine the number of cloud functions used at the runtime.
Furthermore, many of these systems delegate communication and actual offloading to MPI since it provides a portable solution for HPC systems and can benefit from fast networks.
\toolname{} provides its own dispatcher implementation without relying on MPI, which is often unavailable in the cloud and not elastic enough for serverless offloading.

%
%
%

\paragraph{Cloud OpenMP}
%
%
%
%
In OpenMR, Wottrich et al. proposed to offload parallel loops to Apache Hadoop MapReduce runtime in the cloud~\cite{wottrichCloudbasedOpenMPParallelization2014}.
New OpenMP directives are designed to execute MapReduce jobs, with a restricted execution model since OpenMP synchronization directives cannot be supported there.
However, this work does not implement an actual compiler for the extensions and instead conducts manual source code modifications for selected benchmarks.

%
OmpCloud~\cite{yviquelCloudOpenMPOffloading2017,yviquelClusterProgrammingUsing2018} offloads parallel loop into Spark by representing cloud resources as a new device type in OpenMP.
Input data is sent to object storage in the cloud or an HDFS server, and parallel regions are mapped to Spark's map-reduce execution model.
The target code is embedded in the fat binary with host code, and the Java Native Interface (JNI) is needed for compatibility with Scala used in Spark.
Since Spark jobs operate in a distributed setting with data mapped to Resilient Distributed Datasets (RDDs), OpenMP synchronization directives like atomic operations and barriers are not supported.
OmpCloud has been used to offload ray tracing~\cite{mortattiAutomaticRayTracerCloud2018},
a case study similar to ours (Section~\ref{sec:eval-raytracing}).
The operation is perfectly parallelizable since each pixel is rendered independently.
However, the evaluation revealed scalability issues in Spark offloading: non-negligible overheads explained by a bottleneck in the Spark system, and multithreading within a single Spark node that is less efficient than using OpenMP.

These solutions use distributed cloud systems fundamentally different from serverless functions in \toolname{}.
Spark is deployed on a dedicated cluster that cannot be quickly rescaled since allocating a new virtual machine can take up to several minutes.
Even though OmpCloud can start and stop virtual machines automatically,
decreasing startup overheads requires users to pay to keep the cluster ready and available.
Meanwhile, serverless functions do not generate any costs when unused and can be scaled up rapidly, with commercial systems achieving up to 1,000 new execution units every 10 seconds~\cite{awsLambdaScaling}, and open-source systems scaling up to up to 2,500 new functions in a second~\cite{10.1145/3694715.3695966}.
OmpCloud results show significant overhead of the Spark system that is comparable to or even larger than the cost of moving data from host to target~\cite{yviquelCloudOpenMPOffloading2017}, which is the main bottleneck of serverless offloading in \toolname{}.
Furthermore, the low efficiency of parallelization in small benchmarks is explained by the cost of initializing Spark context, and authors propose to employ in the future advanced Spark systems that retain the context alive between jobs~\cite{mortattiAutomaticRayTracerCloud2018}.
This problem is similar to cold starts, and \toolname{} helps to decrease the negative performance effects of cold starts by facilitating C++ programming in serverless (Section~\ref{sec:eval-pi-cold}).

\subsection{GPU}
\label{sec:related_gpu}

Offload was an early approach for programming heterogeneous multicore systems, demonstrated with the Cell Broadband Engine (BE)~\cite{cooperOffloadAutomatingCode2010}.
An offload block could contain a list of variables copied to the accelerator's local context, similar to a C++ lambda function with capture by value.
However, data movement of arrays was handled by explicitly marking pointers as \emph{outer} to trigger access to a software cache or DMA.
Since offloaded functions are compiled to the device only, a source-to-source translator was used to duplicate the call graph and create copies of functions used on the device.
This reduces the developer's effort but creates issues in type inference, function pointers, and handling projects with multiple translation units.

In the field of GPU programming, single-source programming models have been developed to compile C++ code into
accelerators~\cite{Heller_2016,10.1145/2854038.2854041,10.1145/2791321.2791345}.
Early attempts include C++AMP~\cite{gregory2012cpp}, which was supported initially only by the Microsoft compiler and later added to AMD's hcc compiler~\cite{hcccompiler}; both implementations have since been deprecated~\cite{ampdeprecation}.
Offload was an early approach for programming heterogeneous multicore systems~\cite{cooperOffloadAutomatingCode2010}, which extended variable declarations with new to handle data movement and employed call graph duplication to create copies of functions offloaded to the device.
Single-source programming of GPUs requires dedicated compilers and compilation and code generation are simplified at the cost of forcing users to change their source code.
While CUDA requires users to annotate functions compiled into device code and handle data copies, OpenMP offloading to GPUs automatically determines functions compiled to the device and data handling with implicit and explicit data mapping~\cite{antaoOffloadingSupportOpenMP2016}.
Later, the support for OpenMP offloading to GPUs has been added to LLVM~\cite{antaoOffloadingSupportOpenMP2016}.
Different device codes are embedded in a single fat binary to avoid recompilation when switching devices.

SYCL~\cite{reyes2016sycl} is a single-source programming model for GPUs, where programmers use dedicated data accessors to their code and provide explicit kernel naming~\cite{10.1145/3078155.3078187}.
However, functions compiled to the device are determined automatically without explicit annotations.
SYCL has been adopted by many solutions, including Codeplay's ComputeCpp~\cite{computecpp}, Intel oneAPI  Data Parallel C++ (DPC++)~\cite{dpccompiler},
and the AdaptiveCpp~\cite{alpayOnePassBind2023}.
The latter is the only single-pass compiler, which stores the kernel code in an intermediate representation and lowers it later to the target ISA in a JIT fashion.
This improves portability at the cost of a slight decrease in performance on some devices.

\toolname{} takes a different approach as we avoid embedding offloaded code in a single binary and instead create multiple deployments at compilation time.
We do not require portability because the target architecture of serverless functions must be known at the deployment time.
We apply code duplication at the LLVM module level.
Even though \toolname{} does not target GPU devices, it could incorporate similar methods to determine functions used in the serverless target to reduce the size of generated binaries.

\textbf{Remote Devices}
Kasmeridis et al.~\cite{kasmeridisTransparentRemoteOpenMP2024} extended the OpenMP cluster solution~\cite{keftakisExperiencesTaskbasedProgramming2022} with remote offloading.
A new module virtualizes the remote device by forwarding all data and kernels at runtime.
This solution is similar to API remoting that has been previously implemented through hijacking API calls~\cite{5547126} or compiler-assisted translation of API calls~\cite{6495928}.

Patel et.~\cite{patelRemoteOpenMPOffloading2022} proposed offloading OpenMP to remote CPU and GPU targets.
The code is compiled into a single binary, and they adapted the LLVM/OpenMP plugin API to explicitly copy the device's binary code to the remote server.
Users are expected to preallocate execution servers and pass their configuration through an environment variable to the host application.
The implementation can serialize data with protobuf and custom serialization, and it targets clouds with gRPC and HPC clusters with UCX support.
The UCX transport layer was a focus of optimizations~\cite{luEfficientRemoteOpenMP2022} and was later replaced with an MPI communication layer~\cite{shanMPIbasedRemoteOpenMP2023}.

In \toolname{}, the deployment is more complicated as we have to preallocate multiple functions before execution, and functions can come with many external dependencies that are linked dynamically.
Furthermore, runtimes like AWS Lambda require shipping the \emph{libc} with all dependencies, and the deployment package of even simple functions can weigh dozens of megabytes.
In the future, \toolname{} can help OpenMP with offloading to serverless clouds by handling alternative entry points and implementing a virtual OpenMP device that delegates invocation to our runtime.

\subsection{Others}
\label{sec:related_cloud_offload}
gg~\cite{234886} and llama~\cite{llamaserverless} offload UNIX processes tasks to the serverless cloud, with a particular focus on offloading compilation and linking steps through integration into a build system.
These solutions operate on a different granularity of the problem since an entire process is executed on an AWS Lambda function instead of a single C++ function like in \toolname{}.

Finally, the concept of an \emph{alternate} entry point can be found in COBOL.
There, it is used to start the program execution from different places in the code~\cite{cobolguide}.
Thus, COBOL's alternate entry points effectively implement logically separated subprograms~\cite{cobolref}.
In \toolname{}, we use alternative entry points to expose multiple remote targets while maintaining the single source implementation and similar compilation workflow.

\section{Discussion}

\textbf{Why a C++ compiler extension?}
Our primary motivation for supporting automatic offloading in C++ is to leverage the many existing codebases, particularly in scientific computing.
These can be supported without requiring users to rewrite their applications in a different language.
While a compiler patch requires additional effort, this burden will likely decrease once compile-time reflection is adopted in C++26.
This advancement could potentially simplify our solution to a source-to-source translator for implementing alternative entry points.

Languages with built-in metaprogramming capabilities, such as Rust, can support alternative entry points without compiler modifications.
We validate a proof-of-concept implementation that uses Rust's procedural macros to adapt functions for dispatching and offloading, utilizing the \texttt{syn} and \texttt{quote} crates to convert between Rust tokens and syntax tree.
The implementation duplicates the offloaded function: one version (enabled in default compilation mode) replaces the function body with a simple dispatcher that loads a shared library and locates the corresponding symbol; the second version preserves the original function body and is exported as a shared library during a separate compilation.
A Rust-based approach could increase the adoption of compiled languages in serverless environments while still benefiting from \toolname{} infrastructure for deployment, cloud resource management, and efficient runtime dispatch.

\textbf{Does serverless need explicit data management?}
\toolname{} transmits all function data during dispatch and does not provide an interface for explicit memory movement, unlike the \texttt{map} and \texttt{data} directives in OpenMP or memory copy functions in CUDA.
In the current serverless function model, users cannot benefit from asynchronous data movement or maintain warm data on the remote side.
The payload can only be sent once with each invocation request, and consecutive invocations cannot be guaranteed to target the same remote environment.
These operations would be feasible in a stateful serverless model~\cite{254432,10.1145/3698038.3698567}.
There, \toolname{}'s dispatcher module could be extended with operations to support state management and data sharing between functions.

\textbf{\toolname{} versus OpenMP}
The problem of offloading computations has been extensively studied in OpenMP (Section~\ref{sec:related_omp} and~\ref{sec:related_gpu}), and \toolname{} addresses several similar challenges.
OpenMP's offloading solution handles data passing with implicit capture and explicit mapping directives, implements user-defined serialization via mapper directives, uses target directive to annotate code for device offloading, generates unique kernel names, and compiles offloaded kernel entry points by embedding device code in fat binary.
Cppless differs in its deployment approach, requiring separate binaries for each offloaded function and mandatory data serialization.
However, we chose to implement a new model primarily because serverless OpenMP would face significant challenges in data locality, communication, and scalability.

First, applying the OpenMP model to serverless environments could create a false sense of compatibility and lead to poor performance since functions operate in a distributed model.
OpenMP applications are implemented with the assumption of low-latency access to shared memory, which is not the case in a serverless environment.
While OpenMP's offloading capabilities could restrict function address spaces to explicitly mapped data and serverless can implement disjoint address spaces~\cite{jacobExploitingFineCoarseGrained2015}, performance would still be compromised.
A single OpenMP device can share address space among dozens of CPU cores or hundreds of GPU cores, whereas AWS Lambda scales to only six virtual CPUs per function.
Second, the OpenMP standard includes features incompatible with serverless computing, such as synchronization directives.
Implementing these would require suitable cloud storage solutions~\cite{10.1145/3361525.3361535,copik2022faaskeeper}, increasing both cost and complexity.
Finally, \toolname{} can target applications beyond the scientific workloads, such as microservices~\cite{10.1145/3297858.3304013}, where OpenMP is not commonly used.

\subsection{Future Work}
We demonstrate \toolname{} by offloading to AWS Lambda.
Other platforms can be supported even if they lack C++ support, for example, by cross-compiling into WebAssembly deployed with a Node.js shim.
We validated this approach with Clang for Google Cloud Functions, and leave the integration of the toolchain in \toolname{} for future work.
\toolname{} could generate leaner deployments by implementing selective compilation.
There, we can benefit from approaches taken by OpenMP.
Another potential enhancement would be supporting variables captured by reference, using explicit annotation to prevent accidental modifications and issues with dangling references (Section~\ref{sec:implementation_limitations}).

Finally, fault tolerance is currently the full responsibility of the user.
\toolname{} could be extended to catch exceptions thrown inside functions, return failures with appropriate error information, and throw a custom exception on the host side.
When available, the runtime could return the exception's type and the explanation provided by the \texttt{what} function of \texttt{std::exception}.

\section{Conclusions}

We present \toolname{}, a single-source approach to program serverless C++ functions.
\toolname{} overcomes the inherent constraints of C++,
and offers developers a streamlined way to transparently offload tasks to FaaS platforms.
We introduce compiler extensions and meta-programming techniques to transition the runtime code deployment to compilation time.
With a selection of parallel benchmarks, we demonstrate that serverless functions can be effortlessly and efficiently integrated into C++ applications.
In \toolname{}, users employ parallel functions to elastically extend computing resources and accelerate their applications with negligible cost overheads.

\begin{acks}
	This project has received funding from the European Research Council (ERC) under the European
	Union’s Horizon 2020 program (grant agreement PSAP, No. 101002047).
	This work was partially supported by the ETH Future Computing Laboratory (EFCL), financed by a donation from Huawei Technologies.
	We thank Amazon Web Services for support through AWS Cloud Credit for Research program,
	and Swiss National Supercomputing Centre (CSCS) for access to compute resources.
\end{acks}

\bibliographystyle{ACM-Reference-Format}
\bibliography{serverless,cppless}

\end{document}